\documentclass[final,1p,times,authoryear]{elsarticle}

\usepackage{amssymb}
\usepackage{amsthm}
\usepackage{amsmath}
\usepackage{xcolor}

\DeclareMathOperator*{\argmin}{arg\,min}

\usepackage{algorithm}
\usepackage[noend]{algpseudocode}

\usepackage[hidelinks]{hyperref}
\usepackage{subfig}
\usepackage{caption}

\makeatletter
\def\namedlabel#1#2{\begingroup
    #2%
    \def\@currentlabel{#2}%
    \phantomsection\label{#1}\endgroup
}
\makeatother

\DeclareCaptionLabelFormat{cont}{#1~#2\alph{ContinuedFloat}}
\newtheorem{re.}{Remark}
\journal{Journal of Cleaner Production}

\begin{document}

\begin{frontmatter}
\title{Reducing Total Trip Time and Vehicle Emission through Park-and-Ride -- methods and case-study}

\author[a]{Ayane Nakamura}
\author[b]{Fabiana Ferracina}
\author[c]{Naoki Sakata}
\author[d]{Takahiro Noguchi}
\author[e]{Hiroyasu Ando}

\affiliation[a]{organization={Graduate School of Science and Technology},
            addressline={University of Tsukuba}, 
            city={Tsukuba},
            postcode={305-8573}, 
            state={Ibaraki},
            country={Japan}}
\affiliation[b]{organization={Department of Mathematics and Statistics},
            addressline={Washington State University}, 
            city={Vancouver},
            postcode={98686}, 
            state={Washington},
            country={U.S.A.}}
\affiliation[c]{organization={NEC Solution Innovators, Ltd.},
            postcode={136-8627}, 
            state={Tokyo},
            country={Japan}}
\affiliation[d]{organization={Center for Social Innovation in Market Design, Institute for Economic Studies},
            addressline={Keio University}, 
            postcode={108-8375}, 
            state={Tokyo},
            country={Japan}}  
\affiliation[e]{organization={Advanced Institute for Materials Research (AIMR)},
            addressline={Tohoku University}, 
            city={Sendai},
            postcode={980-8577}, 
            state={Miyagi},
            country={Japan}}


\begin{abstract}
This study addresses important issues of traffic congestion and vehicle emissions in urban areas by developing a comprehensive mathematical framework to evaluate Park-and-Ride (PnR) systems. The proposed approach integrates queueing theory and emissions modeling to simultaneously assess waiting times, travel times, and vehicle emissions under various PnR usage scenarios. The methodology employs a novel combination of Monte Carlo simulation and matrix geometric analytic methods to analyze a queueing network representing PnR facilities and road traffic. A case study of Tsukuba, Japan demonstrates the model's applicability, revealing potential reductions in social costs related to total trip time and emissions through optimized PnR policies. Specifically, the study found that implementing optimal bus frequency and capacity policies could reduce total social costs by up to 30\% compared to current conditions. This research contributes to the literature by providing a unified framework for evaluating PnR systems that considers both time and environmental costs, offering valuable insights for urban planners and policymakers seeking to improve transportation sustainability. The proposed model utilizes a single server queue with a deterministic service time and multiple arrival streams to represent traffic flow, incorporating both private cars and public buses. Emissions are calculated using the Methodologies for Estimating Air Pollutant Emissions from Transport (MEET) framework. The social cost of emissions and total trip time (SCETT) is introduced as a comprehensive metric for evaluating PnR system performance.
\end{abstract}


\end{frontmatter}

\section{Introduction}
\label{sec:intro}

Urban transportation systems face mounting challenges from traffic congestion, air pollution, and the need for sustainable mobility solutions. One solution for alleviating such traffic congestion and air pollution is the introduction of a data-driven and parametrized Park-and-Ride (PnR)  system. PnR's are common in the USA and Europe, and are essentially large parking lots with a hub for public transport (e.g., buses or autonomous driving cars). PnR's are specifically located outside of urban centers, giving people in the suburbs access to city centers without the use of their own vehicles. The individual's car or bike is left at the PnR during the day and retrieved upon the owner's return.  Ultimately, PnR's change a city's overall traffic behavior by reducing the number of individual trips going into the city center. This reduces the number of cars on the road, traffic congestion, and road emissions. PnR facilities have emerged as a promising strategy to address the issues of congestion and pollution by encouraging the use of public transit and reducing single-occupancy vehicle trips into city centers \citep{futuretransp1010006}. However, the effectiveness of PnR systems depends on complex interactions between travel behavior, transit operations, and environmental impacts, which is overlooked in most studies.

Travel time and vehicle emissions are influenced by the number of travelers, types of vehicles, speed, and the number of vehicles on the road. Traffic congestion is a common cause of increased travel times and vehicle emissions. Congested traffic (approximately 20 km/hr) not only increases travel time but also significantly raises harmful gas emissions from vehicles. During congested traffic and rush hour, carbon monoxide, hydrocarbons and nitrogen oxide emissions all increase significantly from their levels during smooth traffic flow conditions (approx. 50-80 km/hr) \citep{sjodin1998road, de2000environmental, frey2001emissions}.

However, there is the risk of solving one problem and creating another: the increment of waiting times for customers may not be negligible when we implement a PnR system. If the waiting time becomes too large by promoting people to use public transport, social satisfaction of PnR would decrease even though traffic congestion and emission would be significantly reduced. Therefore, it is important to simultaneously consider all the three elements; waiting time, traveling time, and emission, when we discuss the implementation and optimal operation policy of a PnR system.

This study aims to develop a comprehensive modeling framework to evaluate and optimize PnR systems, addressing a critical gap in the literature by integrating time-based and emissions-based costs within a unified analytical approach. The proposed methodology combines advanced queueing theory with emissions modeling to capture the multifaceted impacts of PnR implementation on urban transportation networks. To evaluate the time cost and environmental impact of PnR fairly and simultaneously, we propose a social cost function that monetarily measures the combined total of time-based and emissions-based costs. As the ultimate goal of this study, we propose an optimization method for PnR system transporters that minimizes social cost and supports system administrator decision-making.

Our success in reaching this goal is demonstrated by the following contributions:
\begin{itemize}
     \item Development of data-driven decision-making process for socially optimal PnR transit policy (parametrized by frequency and capacity of transporters), which can be easily implemented by the system administrator without specific expertise.
    \item Verification of the model and socially optimal decision-making process of PnR via application to a real-world case study in Tsukuba, Japan, emphasizing its practical utility for urban transportation planning and policy analysis.
    \item As a result of the application, reduction of social costs by up to 30\% are achieved under the proposed socially optimal PnR transit policy in Tsukuba city, when compared to current conditions.
    \item Actionable insights from the case study analysis, e.g., potential for further social cost reduction by considering car usage rate as model parameter, are provided and discussed.
\end{itemize}

The findings of this study have significant implications for urban planners, transit agencies, and policymakers seeking to implement or optimize PnR systems. This research supports evidence-based decision making in sustainable urban transportation planning by: (a) quantifying the multi-dimensional trade-offs in PnR system implementation; (b) developing a transferable analytical framework that can be applied across diverse urban contexts; and, (c) providing a social cost function that captures both economic and environmental considerations.

The remainder of this paper is organized as follows. In Section \ref{sec:litrev}, we provide background on the current literature for queueing theory, emission projections and PnR systems. Based on gaps from the literature, we develop our hypotheses and emphasize how this study's contributions fill in the gaps. In Section \ref{sec:method}, we describe our proposed methodology: integration of the queueing and emission models, and introduction of the total social cost for PnR optimization scheme. Next, we demonstrate and validate our approach by performing a case-study of Tsukuba city in Section \ref{sec:case}. A transit policy optimization problem in terms of the social cost given the Tsukuba PnR system is presented, solved and some practical findings are provided in Section \ref{sec:transitpolicy}. Finally, we conclude this paper with some remarks and offer possible avenues for future research in Section \ref{sec:conclusion}.

\section{Literature Review}\label{sec:litrev}

In this section, we provide a research survey on the respective topics of: queueing theory (Subsection \ref{subsec:queue}), emission modeling (Subsection \ref{subsec:emission}), and the PnR system (Subsection \ref{subsec:PnR}), which have been so far considered separately by the current literature. In Subsection \ref{sub:litcont}, we describe the contributions of this study to the current literature in more detail. 

\subsection{Queueing theory}
\label{subsec:queue}

\begin{figure}[H]
\begin{center}
\includegraphics[width=100mm]{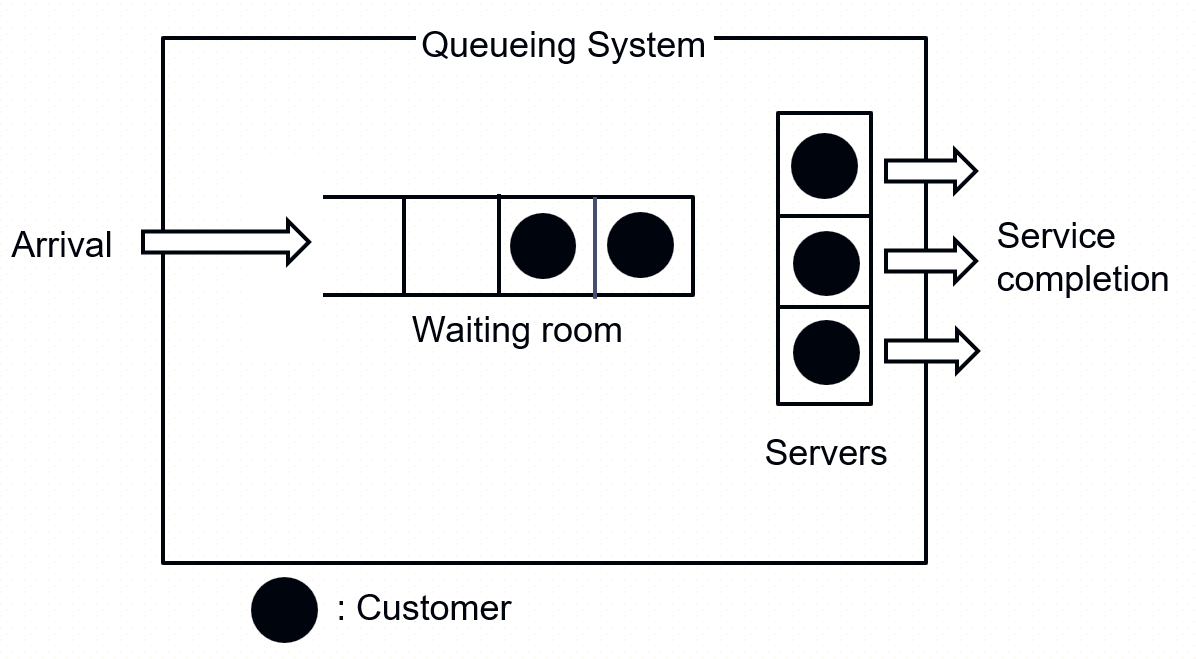}
\caption{Schematic of queueing model.} \label{queueingmodel}
\end{center}
\end{figure}

Queueing theory, which is generally considered as a branch of operations research, is a method of evaluating waiting times for queues or, more generally, the dynamic behavior of customers in various systems.  Examples include people in a queue to be served by a teller at a bank or a cashier at a store, or telecommunication systems such as call-center. A queueing model consists of three elements; a waiting room (sometimes referred to as a buffer), customers, and servers, as seen in Figure \ref{queueingmodel}. The queue itself can be described with Kendall's notation \citep{kendall1953stochastic}, A/S/$c$ queue, where A describes the arrival process, S the service time distribution, and $c$ the number of servers. The most commonly used processes are Markovian (M) (i.e., inter-arrival time or service time are exponentially distributed), Degenerate (D) (i.e., deterministic or fixed inter-arrival time and service time), or General (G), where durations follow an arbitrary distribution. 

Thus, within a system, customers arrive at the waiting room according to some stochastic process A, to receive a service with service time given by S, delivered by some number of servers $c$. Most models operate on a First Come First Served (FCFS) basis, where customers are served in the order of arrival. If a server is available, a customer can receive the service immediately, otherwise they wait in the waiting room, or queue. Once a server is available and the service is completed, the customer leaves the system. In this paper, we present a queueing model that captures the dynamics of customers and vehicles in a PnR system. Here, customers randomly choose between the single service (i.e., private car) and batch service (i.e., public transportation).

The batch (bulk) service queueing models, in which customers receive a common service as a group, were pioneered by \cite{bailey1954queueing}, and have been studied for a long time \citep{sasikala2016bulk}. However, queueing models in which customers probabilistically choose between the single and batch services are part of the recent literature such as the three studies by \cite{math12182820}, \cite{Zhou2024} and \cite{icores2024}. In \cite{math12182820}, they presented the exact and approximated analyses for sojourn time distributions in such model. In \cite{Zhou2024} and \cite{icores2024}, the social optimization or revenue maximization were considered in the context of a queueing game (see \cite{hassin2003queue} for details on queueing game), in which customers behave strategically in a system with options for single or batch services. Economical analysis of such queues often introduces a factor which allows time cost to be converted to monetary cost \citep{hassin2003queue}. Referring to this method, we can consider the social cost which depends on the waiting and traveling times of customers.

As we will later describe in detail, a method for evaluating traffic congestion on roads by means of queueing theory was proposed in \cite{vandaele2000queueing}. Although traffic congestion has a significant impact on vehicle emissions (refer to the detailed survey in Subsection \ref{subsec:emission}), the fusion of the queueing and emission models has not yet been studied until now. Perhaps due to the travelers' transportation choice behavior being sensitive to system parameters \citep{Zhang2012, huang2022analysis}, which would mean that we should not only control for the vehicles on the road, but also the incentives of being on the road with different types of vehicle. Additionally, travel time prediction studies demonstrate the complex interdependencies between various transportation systems \citep{wu2019review, Zhang2019}, thus waiting time for public transport is intricately connected with traffic congestion, and consequently vehicle emissions. This connection suggests that the social cost of traffic should account for wait times. Therefore, to correctly account for emissions, we must consider the following hypothesis:
\begin{enumerate}
   \item[\namedlabel{h1}{\textbf{H1}}] There is a negative correlation between public transportation waiting times and vehicle travel times. 
\end{enumerate}

Regarding a case-study of transportation system in the context of queueing theory, we can refer to a recent study by \cite{Bauer}. In their paper, simulation experiments and discussion for a bus loop in Tsukuba city, Japan, were presented by considering a queueing network of customers and buses.

\subsection{Emission theory}
\label{subsec:emission}
Vehicle emissions represent a significant contributor to air pollution, impacting both human health and the environment. Understanding the complex mechanisms underlying emission formation is crucial for developing effective mitigation strategies and regulatory policies. The formation of vehicle emissions involves intricate chemical reactions occurring within the engine combustion chamber. Key pollutants include hydrocarbons (HC), nitrogen oxides (NO$_x$), carbon monoxide (CO), and particulate matter (PM). Fundamental principles of combustion chemistry govern the production and transformation of these pollutants, influenced by factors such as fuel composition, engine design, and operating conditions \citep{wallington2006automotive}.

Mathematical models serve as essential tools for simulating and predicting vehicle emissions. Mechanistic models, such as chemical kinetics-based simulations, offer detailed insights into emission formation processes. These models utilize complex algorithms to simulate combustion chemistry and pollutant formation, considering parameters such as engine load, speed, and temperature \citep{sharma2016development}. Statistical regression techniques are also commonly employed in emission modeling. Empirical models derive emission factors based on statistical analysis of vehicle emissions data, correlating pollutant concentrations with driving parameters such as vehicle speed, acceleration, and road conditions. While less computationally intensive than mechanistic models, statistical approaches may lack the accuracy and detail of mechanistic simulations, particularly under non-standard driving conditions \citep{franco2008copert}.

The estimation of vehicle emissions involves predicting the quantity and composition of pollutants released into the atmosphere as a result of vehicular activities. Central to emissions modeling is the \textit{emission factor} method, which calculates emissions based on vehicle activity and emission factors (grams of pollutant emitted per unit of activity). The core equation governing emissions estimation is expressed as:

\begin{equation}
E = AF \times EF,
\end{equation}

where \(E\) represents total emissions, \(AF\) denotes the activity factor (e.g., vehicle kilometers traveled), and \(EF\) signifies the emission factor \citep{vanegas2011dynamic}.

Recent advancements in vehicle emission modeling include the integration of machine learning techniques and data-driven approaches. Machine learning algorithms, such as neural networks and support vector machines, offer the potential to improve emission prediction accuracy by leveraging large datasets and complex patterns in emission data \citep{ramos2020machine, aliramezani2022modeling}. However, challenges remain in developing comprehensive emission models that capture the full range of factors influencing emissions, including vehicle technology, traffic conditions, and environmental factors. Furthermore, the validation and calibration of emission models require extensive field measurements and laboratory tests, presenting logistical and resource challenges \citep{smit2010validation, franco2013road}.

The Methodologies for Estimating Air Pollutant Emissions from Transport (MEET) project is a collaborative effort aimed at developing advanced methodologies for estimating air pollutant emissions from transportation sources \citep{demir2014review}. The project involves researchers from various European institutions and focuses on improving the accuracy and reliability of emission estimation models. One application of the MEET project is in the study of traffic congestion and its impact on air quality. By integrating traffic flow models with emission estimation techniques, researchers can assess the environmental consequences of congestion and develop strategies to mitigate its effects. For example, the MEET project has been used to analyze the emission impacts of different traffic management policies, such as congestion pricing and lane restrictions \citep{panis2006modelling, bai2017evaluating}.

The integration of emissions and congestion models offers a comprehensive approach to understanding the environmental impact of vehicular activities. Connected and automated vehicle studies have shown that travel dynamics directly impact emission profiles \citep{wu2019review, obaid2021comprehensive}, while assessments of transportation system vulnerabilities indicate that travel times and emissions are interconnected system characteristics \citep{lin2022transportation}. These arguments lead us to consider the following hypothesis when modeling emissions:
\begin{enumerate}
     \item[\namedlabel{h2}{\textbf{H2}}] There is a positive correlation between vehicle travel times and emission levels, particularly under congested traffic conditions.
\end{enumerate}
Various methodologies have been proposed for achieving an integrated approach to emission modeling. These include:

\textbf{Sequential Modeling:} This approach entails simulating traffic flow first, followed by the calculation of emissions based on the prevailing traffic conditions.

\textbf{Simultaneous Modeling:} Unlike sequential modeling, simultaneous modeling involves concurrently simulating traffic flow and emissions, allowing for feedback between the two models.

\textbf{Integrated Models:} Integrated models merge traffic flow and emissions equations into a unified framework, enabling a comprehensive analysis of their interdependencies.

Recent advancements in integrating emissions and congestion models include dynamic emissions modeling, spatially explicit models, and agent-based modeling \citep{wu2019review}. In this study, we present a novel integrated emissions model where we explore the dynamics of traffic through a park-and-ride system and its interplay with varying transit policies.

\subsection{Park-and-Ride}
\label{subsec:PnR}
Park-and-Ride (PnR) has emerged as an innovative concept wherein individuals park their private vehicles at designated stations and continue their journey using public transportation modes, such as buses or large autonomous vehicles, thus garnering significant attention in contemporary society \citep{futuretransp1010006}.
\begin{description}
\item[\textbf{Overview and Purpose of PnR}:]\mbox{}\\
Originating in the 1920's in the United States, the planning of PnR systems initially operated as independent parking facilities detached from transportation infrastructure. Over time, these systems gained recognition globally, notable examples being in Oxford and Nottingham, UK, Budapest, Hungary, as well as several West-European cities including Munich, Netherlands, and Belgium \citep{PARKHURST2002195, Zijlstra}.

Primarily, PnR systems aim to facilitate access to city centers for work or shopping while alleviating traffic congestion resulting from excessive private vehicle usage \citep{memon2014review, buchari2015transportation}. Additionally, they contribute to mitigating environmental damage associated with increased highway use and parking supply shortages \citep{LIMTANAKOOL2006327, su7079587, mera2023impacts}.

Empirical studies indicate that PnR customers predominantly utilize these services for work and shopping purposes \citep{parkhurst1996economic, Parkhurst, Shirgaokar, Adnan2013EvaluatingTP, adnan2015factors, su13074064}. Factors influencing user incentives include total travel time, income, and trip purpose \citep{Hamid}. Moreover, studies have shown that implementing PnR systems improves job accessibility through reduced travel time \citep{Errol}.
 
\item[\textbf{Location Problem of PnR}:]\mbox{}\\
Several studies have been conducted on optimal PnR location problem, as the location of the facility depends on demand from potential users \citep{Horner}, and the aspects of travel times, costs, and infrastructure \citep{futuretransp1010006}. The survey by \cite{kar2023location} can provide further details on these dependencies.

For instance, the optimal locations for facilities belonging to the PnR system in Delaware City were shown through the analysis of its geographic information system \citep{Faghri}.
\cite{su122310083,sym12081225}, using the multi-criteria method, revealed that the location of a PnR system should be close to the public transport infrastructure, while \cite{HORNER2007255} showed that the closer to railway stations the PnR is the more likely it is to reduce private vehicle use from the transport network. It is also important to take into account the travel cost for customers when determining the locations of PnR \citep{Dickins,LIU2009692}.

\cite{PINEDA201686} proposed an integrated stochastic equilibrium model that considered both private automobile traffic and transit networks to incorporate the interactions between these two modes in terms of travel time and generalized costs. They used their model to show the impact on the travel time incurred by customers of PnR when considering different facilities' locations.
\cite{mesa2022assessing} proposed an optimization procedure to choose PnR facilities' placement on multimodal networks according to different criteria: total travel time including transfers, parking fee and a factor depending on the risk of not having an available spot in the parking facility at the arrival time.
\cite{FARHAN2008445} developed multi-objective spatial optimization model which included three fundamentals of PnR systems: covering as much potential demand as possible, locating park-and-ride facilities as close as possible to major roadways, and siting such facilities in the context of an existing system. This study further showed the validity of their approach for supporting transit planning in in Columbus, Ohio.
Moreover, \cite{Farhan} proposed a discrete linear model for locating PnR facilities as a function of distance, coverage range, and partial regional service in facility siting, and illustrated the flexibility and usefulness of the developed modeling approach for addressing a wide range of planning issues.
\cite{WANG2004709} studied the optimal location and pricing of a PnR facility in a linear city through a deterministic mode choice model, and considered the profit maximization and social cost minimization (i.e.,  minimization of total travel costs) problem in terms of PnR locations. The numerical example illustrated the possibility of a ``win–win'' situation whereby a PnR facility can be profitable, as well as socially beneficial by reducing total travel costs.

In particular, the location of PnR facilities is recommended to 5--6 km from the city center \citep{bolger1992planning}, however the possibility of PnR facilities adding to the traffic congestion should be kept at minimum\citep{spillar1997park}.

\item[\textbf{Demand and Acceptance of PnR}:] \mbox{}\\
Demand of PnR system is influenced by many factors, such as the characteristics of users, prices, or state of road congestion. It is reported the methods for estimating demand become more sophisticated with the added criteria \citep{LAM2007158}, and various studies have been conducted on it. Although our model focuses on the complicated dynamics of customers and vehicles under a given demand, there are several previous studies of demand and acceptance on PnR for readers who are interested, such as: \cite{SEIK1997427,García,Zhi-Chun2007,Stuart2008,KEPAPTSOGLOU201053,Hamer2010,KARAMYCHEV2011455,Zhang2012,Wiseman2012,MINGARDO20137,DU201458,Xianwei,Cao2016,LIU2017162,SONG2017182,Zhang2019,en13133473,webb2020park,Ibrahim,en14010203,su13052644,kumar2021adaptive,yaliniz2022evaluation,huang2022analysis,PITALE2023100065}.

In particular, it should be noted that \cite{SONG2017182} considered the optimization of capacities and transit service frequencies of PnR through an integrated planning framework and showed that the optimal design improves net social benefit, which is the sum of user benefit
and revenues excluding amortized investment and operating
costs of PnR facilities. In our study, we study the social optimization (cost minimization) problem in which the decision variables are the capacities and  frequencies of the public transportation for a PnR system while considering both the time and emission cost.

\item[\textbf{Public Transport with PnR}:] \mbox{}\\
Public transport and PnR system are deeply linked. \cite{Huang2019} verified that PnR system facilities should be located as close as possible to public transport stations, through data analysis of a survey in Melbourne, Australia. Using survey results, \cite{Qin2012} showed that improving the service level of PnR facilities and the comfort for riding bus or railway increase the utilization of PnR facilities in Beijing. 
An empirical case-study of the integration of car parking facilities with public transport in PnR by \cite{CLAYTON2014124} in UK showed that the radial distance to parking, availability of PnR sites in the direction of travel, gender, age, income and party-size become the important factors in a binary logistic regression model. \cite{Mahmoud2014} also presented an empirical investigation on PnR access choices of cross regional commuter in the Greater Toronto and Hamilton Area. They showed that access distance and the station's relative direction (toward the work place) are the primary factors affecting transit choice for PnR users.
In our proposed model, we consider optimizing the operation of public transport via policy in connection with a PnR system, and further adapt this framework to the case-study of Tsukuba city, Japan.

\item[\textbf{Environmental effect of PnR}:] \mbox{}\\
There have been recent studies focusing on the environmental effect of PnR systems, however they do not offer a combined framework which can analyze the waiting time, traveling time, and emission given a PnR design and prescribe optimal transit strategies. 
\cite{rezaei2022park} used a mixed integer linear program in conjunction with a case-study in the US, to reveal the optimal placement of PnR facilities which reduce emission and vehicle kilometer.
Similarly, \cite{xu2023joint} proposed a joint optimization model for PnR facility locations and the alternate traffic restriction scheme,
which optimally designed the alternate traffic restriction areas, the proportion of private cars to be restricted, and the
PnR facility locations. The numerical results in \cite{xu2023joint} showed that the proposed method can minimize the total traveling time and emissions.
Furthermore, \cite{su13094653} showed that the shift to more sustainable transport modes through autonomous electric vehicles and PnR systems significantly reduces pollution in the urban environment.
In particular, \cite{mei2023assessment} came close to developing a framework in their investigation. They assessed the impact of different parking reservation strategies for PnR on vehicle emissions, and developed a framework to reduce vehicle emissions through parking strategy optimization. While they simulated the impact of the PnR system on emissions, they did not consider the impact on total trip times including waiting times or consider the social cost of the different parking strategies.
\end{description}

As described in this section, integration between different pairs of metrics from travel times, waiting times, vehicle emissions and social costs, within a PnR system have been explored. However, several studies mention the limitation of not considering the missing metrics. Their arguments serve to support the hypotheses that further optimization of PnR systems can be achieved by simultaneously considering travel times, waiting times, vehicle emissions and social costs. Thus, we offer the following:
\begin{enumerate}
     \item[\namedlabel{h3}{\textbf{H3}}] The optimization of PnR systems involves a multi-dimensional trade-off between travel times, waiting times, vehicle emissions and social costs.
\end{enumerate}

\begin{figure}
\begin{center}
\includegraphics[width=130mm]{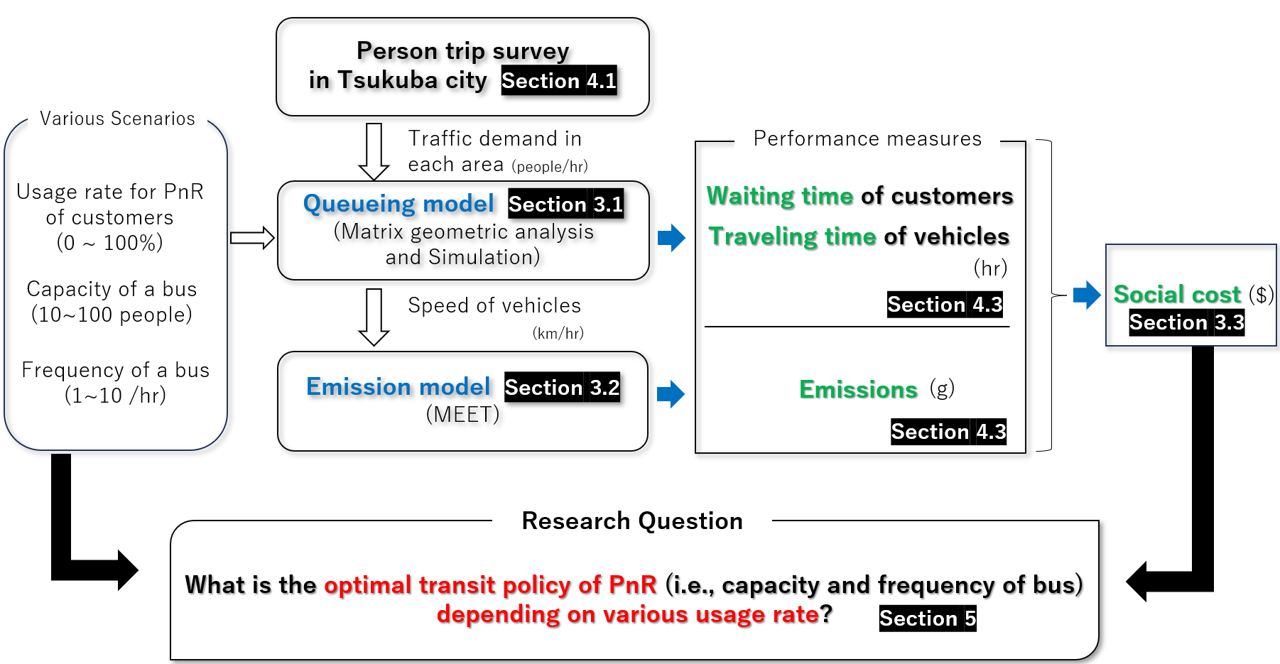}
\caption{Flowchart of the evaluation of Park-and-Ride (PnR) and research question of this paper.} \label{flowchart}
\end{center}
\end{figure}

Several crucial social problems motivate this study. Urban transportation systems are at a critical inflection point, facing unprecedented challenges that demand innovative and holistic solutions. Traffic congestion has become a global urban crisis, with economic losses estimated at billions of dollars annually and significant environmental consequences due to increased carbon emissions. Traditional transportation planning approaches have proven inadequate in addressing the complex interplay between urban efficiency, environmental sustainability and social welfare \citep{LIU2017162, su122310083, mera2023impacts}.

In this study, we present a comprehensive mathematical framework where travel times, waiting times, vehicle emissions, and social cost are modeled and measured together. Through simulations and empirical data, we confirm of the validity of \ref{h1}, \ref{h2} and \ref{h3} using our proposed model. We explore the case-study of Tsukuba, Japan, and realize an optimization scheme for PnR transit policy (based on frequency and capacity of transporters) that minimizes the total social cost. This cost function incorporates both time-based and emissions-based factors, providing a holistic metric for evaluating PnR system performance.

\subsection{Contributions of this paper to the current literature}\label{sub:litcont}
In this study, we propose a comprehensive mathematical framework to evaluate PnR, and apply this method to the case-study of Tsukuba city, in Japan. 
Refer to Fig. \ref{flowchart} for the evaluation of PnR and our research question. The contributions of this paper can be detailed as follows:

\begin{itemize}
    \item \emph{This study is the first exploration in comprehensively considering the social cost on PnR system from the perspective of both time and environmental costs.}
    Aforementioned studies focus on only one of these issues, although both costs are significant and important to society.
    The proposed social cost depends on all the mutually correlated elements, waiting and traveling times endured by productive citizens, and vehicle emissions.
    \item From the methodological perspective, we propose an \emph{integrated mathematical model -- fusion of queueing model and emissions model, and further propose the social cost minimization framework which considers different transit policies.}
    \begin{itemize}
    \item For evaluating the waiting and traveling times of customers under various scenarios of PnR (i.e., the usage rate, the capacity and departure frequency of a bus), we present a queueing model that combines two distinct queues, one capturing the movement of customers and one capturing the movement of vehicles on the road. For this model, we show both the Monte Carlo simulation and the theoretical approximation analysis by means of the matrix geometric method for ensuring the robustness of our findings. The validity of our approach was verified through a comparison between simulation and analytical results.
    \item Moreover, we consider the emission model from Europe's MEET ``Methodologies for estimating air pollutant emissions from transport''. The emission model developed for the MEET project takes as input the speed and type of vehicle, which we provide directly from within the aforementioned queueing model. 
    \item By integrating all the performance measures from the queueing and emission models, we show the total social cost for the PnR system under various scenarios of customer behavior and transportation operation policy. In this total social cost, the time and emission costs are converted to the form of the monetary cost (\$) per capita.
    \item Based on the above total social cost, as our ultimate objective, we develop an optimization framework of transit policy, i.e., the total social cost minimization problem with the capacity and frequency of a bus system as decision variables. Throughout this optimization scheme, we can derive the optimal transit policy aimed at minimizing the social cost for various usage rates of the PnR system.
    \end{itemize}
    \item From the application perspective, \emph{we adapt our method to the case-study of Tsukuba city in Japan where most of the present residents are highly dependent on automobiles.} In this case-study, we geographically divide Tsukuba city into five districts and imagine the central placement of ``hubs'' in which customers park and from which they ride to a conveniently located hub with access to a variety of businesses and transportation to farther places. The travel demands in each districts are estimated by the person trip survey data. Our main findings by the case-study can be summarized as follows:
    \begin{itemize}
        \item The primary findings from the Tsukuba case-study validate our model's robustness and accuracy. By using empirical data from the 2018 Person Trip survey, we were able to simulate total vehicle emissions and trip times, yielding results that closely align with observed data. For instance, our simulation predicted an expected total trip time of 0.4249 hours for trips under 60 minutes, which closely matched the adjusted empirical value of 0.3893 hours. Similarly, our emissions estimates were consistent with real-world measurements, demonstrating that our model can reliably replicate actual conditions.
        \item Moreover, the Tsukuba case-study highlights the potential of our model as a tool for experimenting with and optimizing transit policies. By incorporating real data and producing comparable results, our model can assist urban planners and policymakers in testing various scenarios and interventions to improve transit systems. This capability allows for informed decision-making, aimed at reducing emissions and enhancing travel efficiency, ultimately contributing to the development of more sustainable urban transportation networks.
        \item Specifically, we showed the positive correlation between the traveling time and emissions, and also verified the negative correlation (that is, trade-off relationship) between the waiting time and the traveling time (and accompanying emissions) through the numerical experiments. Our numerical results, under the optimal transit policies such as minimizing the total social cost that considers all the mutually dependent aforementioned measures, showed significant reduction in the total social cost for the current high car usage rate.  Furthermore, it is indicated that much more reduction in the total social cost can be revealed by the less car usage rate compared to the current state -- that implies the social importance of promoting car-free movement.
    \end{itemize}
\end{itemize}

\section{Methods}
\label{sec:method}

The methodological framework developed in this section is designed to systematically investigate and validate the three core hypotheses regarding PnR transportation systems. By constructing an integrated queuing and emissions model, we aim to empirically explore the proposed relationships: the negative correlation between public transportation waiting times and vehicle travel times (\ref{h1}), the positive correlation between vehicle travel times and emission levels (\ref{h2}), and the multi-dimensional trade-offs inherent in PnR system optimization (\ref{h3}). The methodology employs advanced stochastic modeling techniques to capture the complex interactions between transportation system parameters. Through a rigorous examination of vehicle arrival patterns, service times, emission profiles, and social costs, this section describes a robust analytical framework that can quantitatively assess the intricate dynamics of PnR systems.

\subsection{Total trip time model description}
\label{simulationmodel}
\begin{figure}[H]
\begin{center}
\includegraphics[width=120mm]{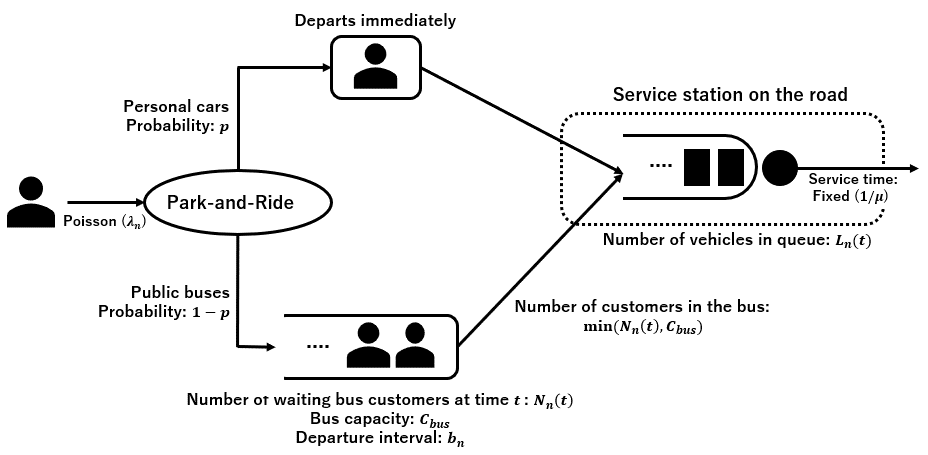}
\caption{Schematic of Park-and-Ride traffic congestion queuing model.} \label{queuemodel}
\end{center}
\end{figure}

In what follows, we describe the Park-and-Ride (PnR) traffic congestion queuing model specifically - see the schematic flowchart of the model in Figure \ref{queuemodel}.
In our model, we assume that a customer arrives at the PnR (spot $n$) according to a Poisson Process with rate $\lambda_n$ and choose either a car, with probability $p$, or a public bus, with probability $1-p$. We set the capacity of a car to one and the capacity of a bus to $C_{bus}$. Customers who choose a private car leave immediately from their PnR spot, whereas public buses leave at a fixed time interval $b_n$. Once a vehicle (i.e., private cars or public buses) departs the PnR, it goes to the city center directly. We assume that the distance between the PnR (at spot $n$) and the city center is $d_n$.

To incorporate traffic congestion into our queueing model, we refer to the study by \cite{vandaele2000queueing}. In this study, the authors assumed that roads are subdivided into segments of length equal to the minimal space needed by one vehicle on the road. Each road segment was considered as a service station, i.e., a queueing model. \cite{vandaele2000queueing} introduced the analytical results under the assumption that the service station were M/M/1, M/G/1, G/G/1 and G/G/1 state dependent queues, respectively. 

They derived the effective speed (i.e., the mean speed for every vehicle) using the existing analytical results of the expected sojourn time in the service station for each assumption. In our model, we assume that the service station is a G/D/1 queue. Specifically, vehicles arrive to the service station by a `general' process, which is a mixed process of the Poisson arrival of private cars, with rate  $p\lambda_n$, and the fixed  inter-departure time of public buses $b_n$. The service time follows a fixed distribution, i.e., a constant $1/\mu$ (where $\mu$ is the service rate). We assume vehicles usually travel at a fixed nominal speed whenever the road is not too crowded. 

Following the theory in \cite{vandaele2000queueing}'s work, the service rate $\mu$ can be calculated as
\begin{equation}
 \label{mu}
    \mu   = v_{NS} \times k_j ,
\end{equation}
where $v_{NS}$ is the nominal speed of cars and $k_j$ (vehicles/distance) the maximum traffic density of the road. The maximum traffic density represents the maximum number of vehicles that can exist on the road per unit distance, implying that $1/k_j$ is the length of a service station.

Let $R$ be  the sojourn time through a vehicle's service station. We can calculate $v$, the speed of the vehicle through the station, as
\begin{equation}
\label{v}
    v = \frac{1/k_j}{R}.
\end{equation}
It should be noted that $v$ is an input parameter of the emission model in Subsection \ref{sec:meet}.
We can then find the traveling time for a vehicle as
\begin{equation}
\label{T}
    T=d_n/v.
\end{equation}
In the approximation model, we let the notation $v$ denote the expected speed of all vehicles, which will be input parameters of the emission model in Subsection \ref{sec:meet}, for simplicity.
There are two stability conditions for this model. The stability condition for the queue of waiting customers for buses is
\begin{equation}
    \label{busstability}
    \lambda_{n} (1-p) <  \frac{C_{bus}}{b_{n}},
\end{equation}
where the left side represents the mean number of arriving customers to the queue for the public bus, and the right side represents for the maximum number of customers served per unit time by public bus. The stability condition for the queue of the service station is
\begin{equation}
    \label{roadstability}
\lambda_{n}  p +  \frac{1}{b_{n}}<  \mu,
\end{equation}
where the left side is the sum of the arrival rate of customers who use private cars and the number of departing buses from a PnR station per unit time (i.e., the total number of vehicles per unit  time), and the right is the service rate $\mu$ of the service station. See the detailed proof for (\ref{busstability}) and (\ref{roadstability}) in \ref{GIM1} and \ref{QBD}, respectively.

$N_n(t)$ is the number of waiting bus customers at time $t$. When the bus departs at time $t^*$ the number of customers in the bus becomes $\min(N_n(t^*),C_{bus})$. For $L_n(t)$, the number of all vehicles (i.e., vehicles which are waiting or receiving the service) in the system of the service station, $\{(L_n(t),N_n(t));t \geqq 0\}$ is a two-dimensional continuous-time stochastic process in the state space $S = \{(i,j);i=0,1,\dots, j=0,1,\dots \}$. Note that we do not have a Markov process due to the assumptions of buses departing at fixed intervals, and the fixed service rate at the service station. 

Under the two stability conditions above, we can numerically obtain the following steady state probabilities:
\begin{equation}
    \pi_{i,j}  =  \lim_{t \to \infty} {\rm P}(L_n(t)=i,N_n(t)=j). \label{eq:ssprobs}
\end{equation}

However, it is not easy to obtain the explicit values for these steady state probabilities for a non Markov process with multiple infinite dimensions. Therefore, we introduce a simple approximation model in the next subsection. We also conduct the simulation experiment for the original model and compare the results with the approximation model.
\begin{figure}[H]
\begin{center}
\includegraphics[width=120mm]{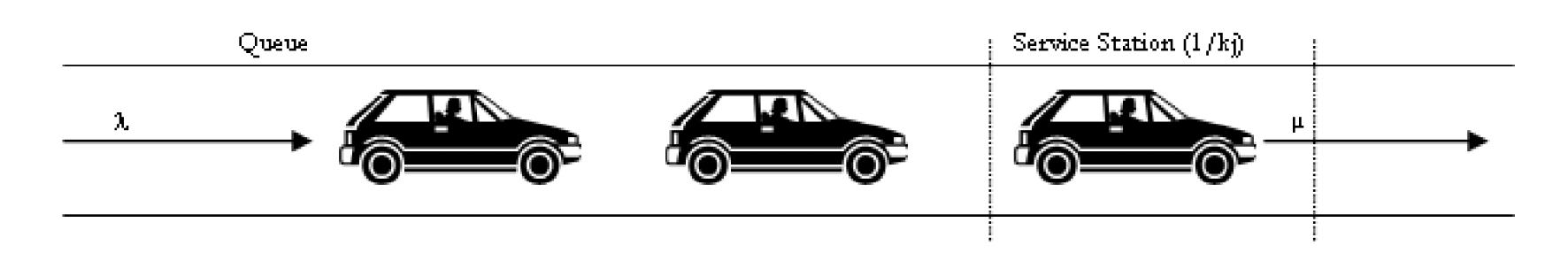}
\caption{Queuing representation of traffic flow \cite{vandaele2000queueing}.} \label{roadmodel}
\end{center}
\end{figure}

\subsubsection{Approximation model}
\label{approximation}
We approximate that the fixed service time of the service station ($1/\mu$) and the fixed inter-departure time of the bus ($b_n$) with Erlang distributions:
\begin{equation}
\label{erlangmu}
{\rm P} \left(X^{(1/\mu)} \leqq x\right) \sim 1 - e^{-l_{q} \mu x} \sum_{n=0}^{l_{q}-1} \frac{(l_{q} \mu x)^n}{n!},
\end{equation}
\begin{equation}
\label{erlangb}
{\rm P} \left(X^{(b_n)}\leqq x\right) \sim 1- e^{-(l_{r}/ b_n) x} \sum_{n=0}^{l_{r}-1} \frac{\{(l_{r}/ b_n) x\}^n}{n!}.
\end{equation}
These Erlang distributions can be interpreted as the sum of $l_q$ $(l_r)$ exponentially distributed random variables of parameters $l_q\mu$ and $l_r/b_n$, respectively.
These distributions converge to the fixed distributions (fixed values $1/\mu$ and $b_n$, respectively) when $l_{q} \rightarrow \infty$ ($l_{r} \rightarrow \infty$). Note that we conduct the numerical experiment under high settings of Erlang phases, i.e., $l_q$ and $l_r$.
The Erlang distribution is a phase-type distribution, therefore this assumption enable us to model the whole system as a continuous-time multidimensional Markov chain. 

Let $L_n^\ast(t)$, $N_n^\ast(t)$, $U_n^\ast(t)$ and $B_n^\ast(t)$ denote the number of all the vehicles in the service station, the number of waiting bus customers, the progress of the Erlang distribution for the service time of the service station and that for the inter-departure time of buses, at time $t$, respectively. It is easy to find that $\left\{(L_n^\ast(t),N_n^\ast(t),U_n^\ast(t),B_n^\ast(t));t \geqq 0 \right\}$ forms a four-dimensional continuous-time Markov chain in the state space $S^\ast$, where $S^\ast$ is defined as
\begin{equation}
    S^\ast = \{(i,j,q,r);i=0,1,\dots,\  j=0,1,\dots,\  q=0,1,\dots,l_{q}-1,\  r=0,1,\dots,l_{r}-1  \}. \label{eq:sstar}
\end{equation}

We can then define the steady state probabilities:
\begin{equation}
    \pi_{i,j,q,r}^\ast  =  \lim_{t \to \infty} {\rm P}(L^\ast_n(t)=i,N^\ast_n(t)=j,U_n^\ast(t)=q,B_n^\ast(t)=r). \label{eq:pistar}
\end{equation}

Note that we do not need to derive the joint probabilities of the number of all vehicles in the service station and the number of waiting bus customers. Our research question is to observe the trade-off between the mean waiting time and the mean traveling time for customers. Hence, it is enough to derive each distribution respectively. Therefore, we can write the steady state probabilities for each distribution as:
\begin{equation}
    \xi_{i,q,r}^\ast  =  \lim_{t \to \infty} {\rm P}(L^\ast_n(t)=i,U_n^\ast(t)=q,B_n^\ast(t)=r) = \sum_{j=0}^{\infty}\pi_{i,j,q,r}^\ast, \label{eq:xistar}
\end{equation}
\begin{equation}
    \omega_{j,r}^\ast  =  \lim_{t \to \infty} {\rm P}(N^\ast_n(t)=j,B_n^\ast(t)=r)=\sum_{i=0}^{\infty}\sum_{q=0}^{l_q-1}\pi_{i,j,q,r}^\ast. \label{eq:omegastar}
\end{equation}

Next we derive the above steady state probabilities ($\xi_{i,q,r}^\ast$ and $\omega_{j,r}^\ast$) respectively, and calculate each performance measure, following the matrix analytic method (refer e.g., \cite{adan2017analysis} and \cite{latouche1999introduction}).

\subsubsection{Analysis of the service station on the road}
First, we focus on the derivation of $\xi_{i,q,r}^\ast$. We have that $\left\{(L_n^\ast(t),U_n^\ast(t),B_n^\ast(t));t \geqq 0 \right\}$ forms a three-dimensional continuous-time Markov chain under the state space $S_L^\ast := \{(i,q,r);i=0,1,\dots,\   q=0,1,\dots,l_q-1,\  r=0,1,\dots,l_r-1\}$. Our Markov chain forms a Quasi Birth-and-Death (QBD) process, where $L_n^\ast(t)$ is the level and $\left\{(U_n^\ast(t),B_n^\ast(t)) \right\}$ is the phase.

The detailed analysis is given in \ref{QBD}, and we confirm that the stability condition is equivalent to (\ref{roadstability}).

Based on the above, we can obtain several (approximated) performance measures. Let ${\rm E}[L^{\ast}_n]$, ${\rm E}[R^{\ast}_n]$ and ${\rm E}[T^{\ast}_n]$ denote the mean number of vehicles in the service station, the mean sojourn time in the service station, and the mean traveling time of vehicles from PnR (spot $n$) to the city center, respectively. For ${\rm E}[L^{\ast}_n]$, the following holds:
\begin{equation}
    \begin{split}
        {\rm E}[L^{\ast}_n] &= \sum_{i=1}^{\infty} i \mathbf{\xi}_i^\ast \mathbf{e}\\
        &= \mathbf{\xi}_1^\ast  (\mathbf{I} - \mathbf{R}^{(L)})^{-2}\mathbf{e}.
    \end{split}
\end{equation}
We also obtain ${\rm E}[R^{\ast}_n]$ by Little's law \citep{little1961proof}:
\begin{equation}
\label{ER}
    \begin{split}
        {\rm E}[R^{\ast}_n] = \frac{\mathbf{\xi}_1^\ast  (\mathbf{I} - \mathbf{R}^{(L)})^{-2}\mathbf{e}}{p \lambda_n + 1/b_n}.
    \end{split}
\end{equation}
Applying the relationship of (\ref{v}) and (\ref{T}), we obtain
\begin{equation}
\label{ET}
    {\rm E}[T^{\ast}_n] = d_n k_j {\rm E}[R^{\ast}_n].
\end{equation}
Note that $\mathbf{e}$, $\mathbf{I}$ and $\mathbf{R}^{(L)}$ are a vertical vector, an identity matrix and the rate matrix, defined in \ref{QBD}.

\subsubsection{Analysis of waiting bus customers}
Next, we derive $\omega_{j,r}^{\ast}$.
Similar to our prior analysis, we can find that $\left\{(N_n^\ast(t),B_n^\ast(t));t \geqq 0 \right\}$ forms a two-dimensional continuous-time Markov chain under the state space $S_N^\ast := \{(j,r);j=0,1,\dots,\  r=0,1,\dots,l_r-1\}$. Our Markov chain is of GI/M/1-type, where $N_n^\ast(t)$ is the level and $B_n^\ast(t)$ is the phase.
The analysis and the proof for the stability condition, which is equivalent to (\ref{busstability}), are given in \ref{GIM1}.

Using the steady state probabilities and the rate matrix $\mathbf{R}^{(N)}$, we obtain the mean number of waiting bus customers and the mean waiting time at PnR (spot $n$) (denoted by ${\rm E}[N^{\ast}_n]$ and ${\rm E}[W^{\ast}_n]$, respectively):
\begin{equation}
\begin{split}
{\rm E}[N^{\ast}_n] &= \sum_{j=1}^\infty j \mathbf{\omega}_{j}^\ast \mathbf{e}\\
&= \mathbf{\omega}_{1}^\ast (\mathbf{I} - \mathbf{R}^{(N)})^{-2}\mathbf{e},
\end{split}
\end{equation}
and by Little's law,
\begin{equation}
    {\rm E}[W^{\ast}_n] = \dfrac{ \mathbf{\omega}_{1}^\ast (\mathbf{I} - \mathbf{R}^{(N)})^{-2}\mathbf{e}}{(1-p)\lambda_n}.
\end{equation}

Finally, we obtain the mean total trip time of vehicles from PnR (spot $n$) to the city center  (i.e., the sum of waiting time and traveling time for vehicles), ${\rm E}[TotalTrip^{\ast}_{n}]$, by combining the results of two analysis:
\begin{equation}
    \begin{split}
        {\rm E}[TotalTrip^{\ast}_{n}] = {\rm E}[T^{\ast}_{n}] + (1-p){\rm E}[W^{\ast}_{n}]. \label{eq:totaltrip}
    \end{split}
\end{equation}
Refer to Table \ref{tab:params} for a list of the parameters and main performance measures described above. 
\begin{table}[ht]
\caption{Parameters and performance measures used in the approximation model. \label{tab:params}}
  \centering
  \begin{tabular}{lcr}
  \hline
    Parameters  (unit)  & Explanations    \\
    \hline \hline
    $p$  &  Probability that customers use private cars.  \\
    \hline
    $1/\mu$ (vehicle/hr) & Constant service time of service station. \\
    \hline
    $v_{NS}$ (km/hr)  & Nominal speed of vehicles.   \\
    \hline
    $k_j $  (vehicle/km)&  Maximum traffic density. \\
    \hline
    $\lambda_n$ (people/hr)  & Arrival rate of customers to PnR (spot $n$). \\
     \hline
    $C_{bus}$ (people) & Capacity of a bus.   \\
    \hline
    $b_n$ (hr)  & Inter-arrival time of buses at PnR (spot $n$).  \\
    \hline
    $d_n$ (km) & Distance traveled from PnR (spot $n$) to city center.\\
    \hline
    $l_q$ & Erlang phase for $1/\mu$ in (\ref{erlangmu})\\
    \hline
    $l_\mu$  & Erlang phase for $b_n$ in (\ref{erlangb})\\
    \hline
    $\epsilon$  & Extreme small parameter to determine $\mathbf{R}^{(L)}$ and $\mathbf{R}^{(N)}$\\
    &(see Appendices \ref{QBD} and \ref{GIM1}) \\
    \hline

    \\ 
    \hline
    Performance Measures (unit) & Explanations    \\
    \hline
    \hline
    $v$ (km/hr) & Mean speed of vehicles.  \\
    \hline
    ${\rm E}[W^{\ast}_{n}]$ (hr)  & Mean waiting time of bus customers at PnR (spot $n$).  \\
    \hline
    ${\rm E}[T^{\ast}_{n}]$ (hr)  & Mean traveling time of cars from PnR (spot $n$) \\ & to the city center.  \\
    \hline
    ${\rm E}[TotalTrip^{\ast}_{n}]$ (hr)  & Mean total trip time of vehicles from PnR (spot $n$) \\ & to the city center.  \\
     \hline
  \end{tabular}
\end{table}

\subsubsection{Approximation method for estimation of maximum traffic density}
\label{subsec:approxkj}
We propose an approximation method for estimating the maximum traffic density, $k_j$,  as it is challenging to calculate this directly from the data. By assuming that the current traffic behavior in Tsukuba follows a simple M/D/1 queue, which is more tractable than the original G/D/1 queue assumption, we can analytically calculate $k_j$. This calculation utilizes the arrival rate of all vehicles, $\lambda_{current}^{all}$ and the service rate $\mu$. 
In our implementation we assume that the arrival rate of all vehicles is: $\lambda_np + 1/b_n$, while $\mu$ can be computed from the initial parameters.
According to the existing theory (see, e.g., Chapter 6 of \cite{medhi2002stochastic}), one can use $\lambda_{current}^{all}$ and the equation (\ref{mu}), to calculate the approximated mean sojourn time in the service station ${\rm E}[R_{current}]$,
\begin{equation}
     {\rm E}[R_{current}] = \dfrac{1}{\mu} +  \dfrac{\lambda_{current}^{all}/\mu}{2 \mu (1- \lambda_{current}^{all}/\mu)}
        = \dfrac{2 v_{NS} k_j - \lambda_{current}^{all}}{2 v_{NS}^2 k_j^2 - 2 v_{NS} k_j \lambda_{current}^{all}}. \label{eq:ercurr}
\end{equation}
      
Considering the mean traveling time of vehicles, ${\rm E}[T_{current}]$ and using the relationships (\ref{v}) and (\ref{T}), we can obtain an expression for $k_j$,
\begin{equation}
\begin{split}
    {\rm E}[T_{current}] = d_n k_j  {\rm E}[R_{current}] 
    = \dfrac{2 v_{NS} d_n k_j - \lambda_{current}^{all} d_n}{2 v_{NS}^2 k_j - 2 v_{NS} \lambda_{current}^{all}}\\
\implies
    k_j = \dfrac{\lambda_{current}^{all}(2 {\rm E}[T_{current}] v_{NS} - d_n)}{2 v_{NS} ({\rm E}[T_{current}] v_{NS} -d_n)},  \label{eq:kj}
\end{split}
\end{equation}
where ${\rm E}[T_{current}]$ can be calculated by Person Trip Survey \citep{tokyo2018pt} which will be described in detail in Section \ref{sec:case}.

Note that this expression requires a positive denominator. If this is not the case, we simply set $k_j =\lambda_{current}^{all}/v_{NS}$, i.e., assume that every car travels on the road at $v_{NS}$.
\begin{equation}
    \begin{split}
        k_j :=
            \left\{
                \begin{array}{ll}
                    \dfrac{\lambda_{current}^{all}(2 {\rm E}[T_{current}] v_{NS} - d_n)}{2 v_{NS} ({\rm E}[T_{current}] v_{NS} -d_n)}, & {\rm E}[T_{current}] v_{NS} -d_n > 0, \\
                    \dfrac{\lambda_{current}^{all}}{v_{NS}}, & {\rm E}[T_{current}] v_{NS} -d_n \leqq 0.
                \end{array}
            \right. \label{eq:kjid}
    \end{split}
\end{equation}

This $k_j$ is used to obtain the service rate $\mu$ of the service station as in (\ref{mu}). This approximation is justified because of the proposed PnR traffic model, a G/D/1 queue, as the ratio of private cars among all the vehicles is high and the number of public buses is low, making the effect of the non-Poisson arrival process of public buses marginal. In fact, we confirm the validity of this approximation method in comparison with the Monte Carlo simulation in Subsection \ref{sec:results}.

\subsubsection{Setting of numerical experiments} \label{subsec:setting}
We conduct the Monte Carlo simulation of the original model, given in Section \ref{simulationmodel}, and the numerical experiment for the approximation model given in Sections \ref{approximation} and \ref{numerical}.
The main differences between the simulation and approximation model are:
\begin{itemize}
\item In the Monte Carlo simulation, we obtain the sojourn time in the service station and the speed of each vehicle, respectively, and calculate each travel time using (\ref{v}) and (\ref{T}). On the other hand, in the approximation model, we calculate the expected sojourn time (\ref{ER}) and by using this, obtain the expected travel time (\ref{ET}) and expected speed of vehicles by (\ref{v}).
\item In the approximation model, we use the Erlang approximation for the fixed service time of the service station and the fixed inter-departure time of the bus as in (\ref{erlangmu}) and (\ref{erlangb}).
\end{itemize}

The Monte Carlo simulation keeps track of each vehicle dynamic and obtains accurate values for performance measures. Compared to this, the approximation model can obtain only the expected performance measures for all the vehicles, making the results relatively rough. However, the approximation model has an advantage in its computation cost, enabling easy experimentation with various scenarios. We confirm the validity of the approximation model first in Section \ref{numerical} and conduct thorough experiments for our model.

Several assumptions for input parameters $l_q$, $l_r$ and $\epsilon$ were made.
Regarding the Erlang phases, we set $l_q=20$ and $l_r=200$ when comparing the approximation model to the Monte Carlo simulation. The later phase, pertains to the bus arrival distribution which is deterministic and thus requires higher phase for better approximation results. However, when experimenting with different bus policy scenarios, we set $l_r=20$ for computational efficiency. Finally, we set $\epsilon=10^{-10}$.

\subsection{Vehicle emissions model description} \label{sec:meet}
According to \cite{demir2014review}, factors influencing fuel consumption can be divided into five categories: vehicle, environment, traffic, driver and operations. Due to the nature of our study, we focus primarily on the ``traffic'' category, as it accounts for both speed and congestion, and secondarily on the ``vehicle'' category, as it accounts for vehicle and fuel types. As previously explained, we simulate different types of vehicles with different types of fuel flowing through a road queuing system. From this simulation, we can obtain expected travel times and speed per vehicle, which are indicators of traffic congestion. These vehicle qualities and quantities become the parameters in the emissions model we have adopted from the ``\textbf{M}ethodologies for \textbf{E}stimating air pollutant \textbf{E}missions from \textbf{T}ransport,'' or MEET \citep{hickman1999methodology}.

MEET established a classification for vehicle types so the emission factors can be set appropriately within their vehicle category. For our study, we consider the following categories: 
\begin{itemize}
    \item Private cars: Vehicles used for the carriage of passengers and comprising not more than eight seats in addition to the driver's seat. Different emissions factors - i.e. $\eta$ in (\ref{eq:ehot}), apply to either gasoline (see Table \ref{table:gaspc}) or diesel (see Table \ref{table:dieselpc}) type vehicles.
    \item Small buses (Capacity 30 seats): Vehicles used for the carriage of passengers and comprising more than eight seats in addition to the driver's seat, and having a maximum weight exceeding 3.5 tonnes. See Table \ref{table:caplt30} for emissions factors - i.e. $\varepsilon$ in (\ref{eq:hdv}).
    \item Medium buses (Capacity 60 seats): Vehicles used for the carriage of passengers and comprising more than eight seats in addition to the driver's seat, and having a maximum weight exceeding 7.5 tonnes. See Table \ref{table:caplt60} for emissions factors - i.e. $\varepsilon$ in (\ref{eq:hdv}).
    \item Large buses (Capacity 100 seats): Vehicles used for the carriage of passengers and comprising more than eight seats in addition to the driver's seat, and having a maximum weight exceeding 16 tonnes. See Table \ref{table:capub} for emissions factors - i.e. $\varepsilon$ in (\ref{eq:hdv}).
\end{itemize}

In automotive engineering, ``hot emissions'' usually refer to pollutants emitted from the vehicle's exhaust system when the engine is running at operating temperatures. These emissions primarily consist of nitrogen oxides (NOx), carbon monoxide (CO), hydrocarbons (HC), and particulate matter (PM). They are called hot emissions because they are generated when the engine and catalytic converter reach their optimal operating temperatures during normal driving conditions \citep{stone2004automotive}. In MEET, these emissions are calculated using the formula:
\begin{equation}
    E_{hot} = \eta \times m, \label{eq:ehot}
\end{equation}
where $E_{hot}$ is the emission in units of mass/time, $\eta$ the hot emission factor in g/km, and $m$ is the distance traveled per unit time. Emission factors will primarily vary with different vehicle types and speeds, which may also vary with the road system (e.g. urban versus rural) and distances being considered. 

Although simplified, our model can account for vehicle types and speeds, and by simulating each vehicle on the road we can apply MEET's equation for hot emissions estimation. Their complete formula is
\begin{equation}
    E_k = \sum\limits_{i=1}^{i=\mathrm{categories}} n_i \times l_i \times \sum\limits_{j=1}^{j=\mathrm{roadtypes}} p_{i,j} \times \eta_{i, j, k}. \label{eq:ek}
\end{equation}
$k$ indicates the pollutant (e.g. NOx, CO, CO$_2$). $i$ is the number of vehicle categories and $n_i$ is the number of vehicles in category $i$. Since our simulation considers each vehicle individually, we fix $i$ in its category and $n_i = 1$. $l_i$ pertains to the average distance traveled by vehicle of type $i$, while $p_{i,j}$ pertains to the proportion of that distance traveled by vehicle of type $i$ on road type $j$. Since our model considers each vehicle independently and only one type of road, $l_i$ is simply the average distance the model provides and we set $j=1$ and $p_{i,j}=1$. 

$\eta_{i,j,k}$ is the emission factor of pollutant $k$ given vehicle and road types, which we set based on Tables \ref{table:gaspc}, \ref{table:dieselpc}, \ref{table:caplt30}, \ref{table:caplt60} and \ref{table:capub} as appropriate. These tables list the emission factors for different pollutants given vehicle types which conform to the European emission regulation and standards, which although strict, adopting them into our model will not forsake generalizability to the rest of the world. The European Union is a big importer of foreign vehicles, as well as contains several automobile manufactures exporting their vehicles \citep{crippa2016eu}.

Note that tables pertaining to private cars (Tables \ref{table:gaspc} and \ref{table:dieselpc}) describe the formula for the emission factor of each pollutant directly, while the tables pertaining to the heavier vehicles list coefficients. Heavier vehicles have different types of engines depending on their weight limits, however these differences can be accounted with using different coefficients into one emission factor function
\begin{equation}
    \varepsilon(v) = K + av + bv^2 + cv^3 + \frac{d}{v} + \frac{e}{v^2} + \frac{f}{v^3}, \label{eq:hdv}
\end{equation}
where $\varepsilon$ is the rate of emission in g/km for an unloaded goods vehicle, or for a bus or coach
carrying a mean load, on a road with a gradient of 0\%, $K$ is a constant, $a$ through $f$ are coefficients, and $v$ is the mean velocity of the vehicle in km/h \citep{hickman1999methodology}.
Recall that $v$ in our PnR model can be calculated by substituting $k_i$ in (\ref{eq:kjid}) into (\ref{v}) in Subsection \ref{simulationmodel}.
For the assumption of a gradient of 0\%, we consider it is reasonable in Tsukuba city, which is the location of our study, because the most part of the city is a flat terrain \citep{Tsukubacity}.

\subsection{Total social cost model description} \label{subsec:totalcost}
We provide our own model for computing the total social cost when considering total trip mean time and total vehicle mean emissions based on the methods in Subsections \ref{simulationmodel} and \ref{sec:meet}. We will only consider CO$_2$ emissions in our social cost model, as research on the social cost of carbon (or SCC, where carbon refers to CO$_2$) is more readily available. The total \textbf{S}ocial \textbf{C}ost of \textbf{E}missions and total \textbf{T}rip \textbf{T}ime (SCETT), in international dollars (int'l \$), is computed as follows:

\begin{equation}
    \mathrm{SCETT} = \sigma^{\{F, R\}}_{r} \times E^{CO_2}_{(veh, sr, |I|)} + \pi_{r} \times |I| \times \Delta T_{sr}, \label{eq:scett}
\end{equation}
where:
\begin{itemize}
    \item SCETT is the resulting cost in international dollars per capita,
    \item $\sigma^{\{F, R\}}_{r}$ is the social cost of carbon dioxide in international dollar per grams for a region $r$ under either the FUND ($F$) or RICE ($R$) economic models (see Table \ref{tab:scc}),
    \item $E^{CO_2}_{(veh, sr, |I|)}$ is the mean CO$_2$ in grams emitted by all vehicles $veh$, in sub-region $sr$ for time interval length $|I|$, which can be calculated by (\ref{eq:hdv}),
    \item $\pi_{r}$ is the social cost of travel time in international dollar per hour for region $r$. Time spent waiting for transportation and traffic affect social productivity (see Figure \ref{fig:prod}),
    \item $|I|$ is the length of the time interval being considered in hours, this parameter provides flexibility to use emissions and travel time data that have been aggregated into an interval other than one hour,
    \item $\Delta T_{sr}$ is average total trip time (sum of waiting and traveling times) in hours for a sub-region $sr$, which can be calculated by (\ref{eq:totaltrip}).
\end{itemize}

Typically, the estimation of SCC aggregates the impacts of climate change across societies in varying stages of development. \cite{anthoff2019inequality} proposed a SCC model that enabled nuanced consideration of income distribution in different regions of the world by expanding on two prominent tools for computing SCC in the field of climate economics: FUND and RICE. Our total social cost model incorporates the cost per metric ton estimated in \cite{anthoff2019inequality}'s work for both of these tools. Table \ref{tab:scc} provides the SCC computed in their paper for various regions of the world. 

FUND (Framework for Uncertainty, Negotiation, and Distribution), developed by \cite{anthoff2009risk}, is an integrated assessment model designed to evaluate the economic impacts of climate change policies by estimating the costs and benefits associated with mitigation and adaptation strategies. It considers various sectors and regions globally and incorporates uncertainties in key parameters to provide robust policy insights. RICE (Regional Integrated Climate-Economy), developed by \cite{nordhaus2010economic}, is a regional integrated assessment model specifically tailored to analyze the interactions between climate change and the economy at the regional level. It accounts for different regions' economic structures, energy systems, and climate impacts, allowing for a detailed assessment of climate policies' effectiveness and their economic implications. These models serve as valuable tools for policymakers and researchers to assess the societal costs of carbon emissions and inform climate policy decision-making.

When it comes to computing the social cost for total trip time, we need to consider commuting times as opportunity costs. If one's productivity contributes $x$ international dollars per hour to the gross domestic product (GDP) of a region, then if $T$ hours is the total time one spends waiting for a bus or stuck in traffic, that region's GDP does not gain $xT$ international dollars. Figure \ref{fig:prod} shows productivity trends in international dollars per hour from 1980-2019 for eight countries (\cite{feenstra2015next} with major processing by \cite{ourworld2024data}). In our model, we incorporate the social productivity figure as a cost that is multiplied by the average travel time of an individual.

With SCETT, we augment the PnR system to serve as a tool for transportation policy discovery. We incorporate SCETT into our model as follows: for optimal parameters of interest representing PnR hubs, i.e. percentage car usage, bus frequency interval, and bus capacity, we obtain SCETT's argmin of a subset of these optimal parameters while summing over the remaining. For instance, a bus company may gain valuable insights regarding policy decisions by considering what bus frequency interval and capacity would cost society the least. For that purpose, policy managers would like to find the bus frequency and capacity policy constrained by hub $h$ and \% car use $p_{car} = p$:
\begin{equation}
    \argmin\limits_{(b,C)}^{h,p} \left(\sum\limits_{d(h,p)} \sum\limits_{i(h,p)} \left( \sigma^{\{F, R\}}_{reg} \times E^{CO_2}_{(bus+car, h, 4)}(p, b, C) + \pi_{r} \times 4 \times \Delta T_{h}(p, b, C) \right) \right). \label{eq:argminbC}
\end{equation}

In (\ref{eq:argminbC}), $b, C$ represent the bus frequency interval and bus capacity respectively. $d(h,p)$ represents the commute direction (from or to city center) constrained by hub $h$ and \% car use $p$. $i(h,p)$ is one of six 4-hour time buckets constrained by hub $h$ and \% car use $p$. Note that in our current PnR system we only account for emissions of buses and private cars, and the length of time interval $|I| = 4$. The remaining parameters are as described in (\ref{eq:scett}).

We note that the solution $(b, C)$ given by \ref{eq:argminbC} represents the daily bus frequency and capacity (respectively) which minimize SCETT. It is possible to consider this optimization problem more finely and obtain $(b,C)$ on a per hour or time bucket basis, however in this study we demonstrate the model pragmatically and obtain results on a per day basis.

\begin{re.} (Definition of the total social cost):
Our proposed total social cost, SCETT, does not depend on public bus fees or private cars' maintenance costs for customers. The definition used in this study assumes that those monetary costs are paid to third-party entities in society (e.g., bus companies, car companies, and gas stations). Thus, as a social cost, the fees or the maintenance costs are offset and have no impact (for instance, the definition of social welfare in economical analysis for queueing models of Chapter 2 in \cite{hassin2003queue} also does not include the fees). On the other hand, no entity can benefit from the time and emission costs if they are incurred -- these just lower the level of satisfaction for customers and worsen the environmental impact in society. Therefore, it is highly important to discuss both time and emission costs simultaneously through SCETT while considering the balance of both factors.
\end{re.}

\section{Case-study for a PnR system in Tsukuba}
\label{sec:case}
For our case-study, we decided to focus on the Tsukuba area - the Science City. Located in the Ibaraki prefecture of Japan, with an estimated population of approximately 245,000, Tsukuba is just north of the Tokyo metropolitan area. It has one of the largest universities in Japan and it is also the home to several private and public research institutes such as the National Institute of Advanced Industrial Science and Technology (AIST), the Japan Aerospace Exploration Agency (JAXA), the Geospatial Information Authority of Japan (GSI), the National Research Institute for Earth Science and Disaster Prevention (NIED) and the Meteorological Research Institute (\cite{tsukuba2024wiki}  and \cite{japan2020exp}).
In 2022, Tsukuba city was designated as Super Science City by the national government \citep{UnivTsukuba}. 
As a part of Super Science City project, Tsukuba city aims to actualize safe, secure, and comfortable travel in a suburban areas that are highly dependent on automobiles, by
enhancing public transportation services using mobility data. \citep{IbarakiPrefecture}

\subsection{The Person Trip Survey}
The dataset used for the case-study is from the Person Trip Survey \citep{tokyo2018pt} conducted by a council overseeing the Tokyo metropolitan area which includes Tokyo, Kanagawa, Saitama, Chiba, and southern Ibaraki. The Tokyo metropolitan area is the most populous metropolitan area in the world. As of 2017, it is estimated that approximately 38 million people live in this region - as such it is an important region for economy and culture. In order to ensure that the flow of people and goods is supported by the available transportation services and infrastructure, prefectures of the Tokyo metropolitan are established the ``Tokyo Metropolitan Area Transportation Planning Council.'' 

Established in 1968, the council promotes comprehensive urban transportation planning in the Tokyo metropolitan area. As such, it conducts survey studies so that better understanding of person trips and good flows can lead to improvements in transportation services and infrastructure. In particular, the person trip survey investigates the who, when and why of transit in the Tokyo metropolitan area - each resident is given a survey that asks about their movement, such as where they went and how they got there, on the day they fill out the survey.

The survey used in this paper is the 6th person trip survey conducted from September to November of 2018. From 268 municipalities, 630,000 households were randomly selected from 18 million. Questionnaires were mailed to each household and could be completed via written form or online. Table \ref{pt_data} details the features in the person trip survey dataset. For each household, all members of the family filled out the survey and answered questions about each segment of their trip to and from a specific location or various sequential locations.

Out of the randomly selected 630,000 households, 164,934 returned the survey indicating a 26\% household response rate. After the surveys were returned, all responses were aggregated into one dataset of 693,083 observations such that each observation represents one trip segment to a specific location for one member of each household. The number of individuals represented in the responses were 382,667, who may have recorded one-way trips, round trips or no trips. Out of the 382,667 household members represented by the survey, 48,484 of them recorded no trips. Thus the survey depicts the movements of 334,183 members of the Tokyo metropolitan area on an arbitrary workday who in total took 618,769 trips.

Since these observations represent about 1\% of the total movement within the Tokyo metropolitan area, each observation was given a ``magnification factor'' which extrapolates the number of like trips. This extrapolation takes into account each region's:  resident population by sex and age, resident population by the number of household members, workplace population, commuting population, and number of cars owned per household. The expanded estimates allows the survey responses to provide us approximately 74 million trips on an arbitrary workday, which is sufficient to represent the area's 38 million individuals \citep{tpt2018dataguide}.

\subsection{Tsukuba's Park-and-Ride System}\label{numerical}
\begin{figure}
    \centering
    \includegraphics[width=0.64\textwidth]{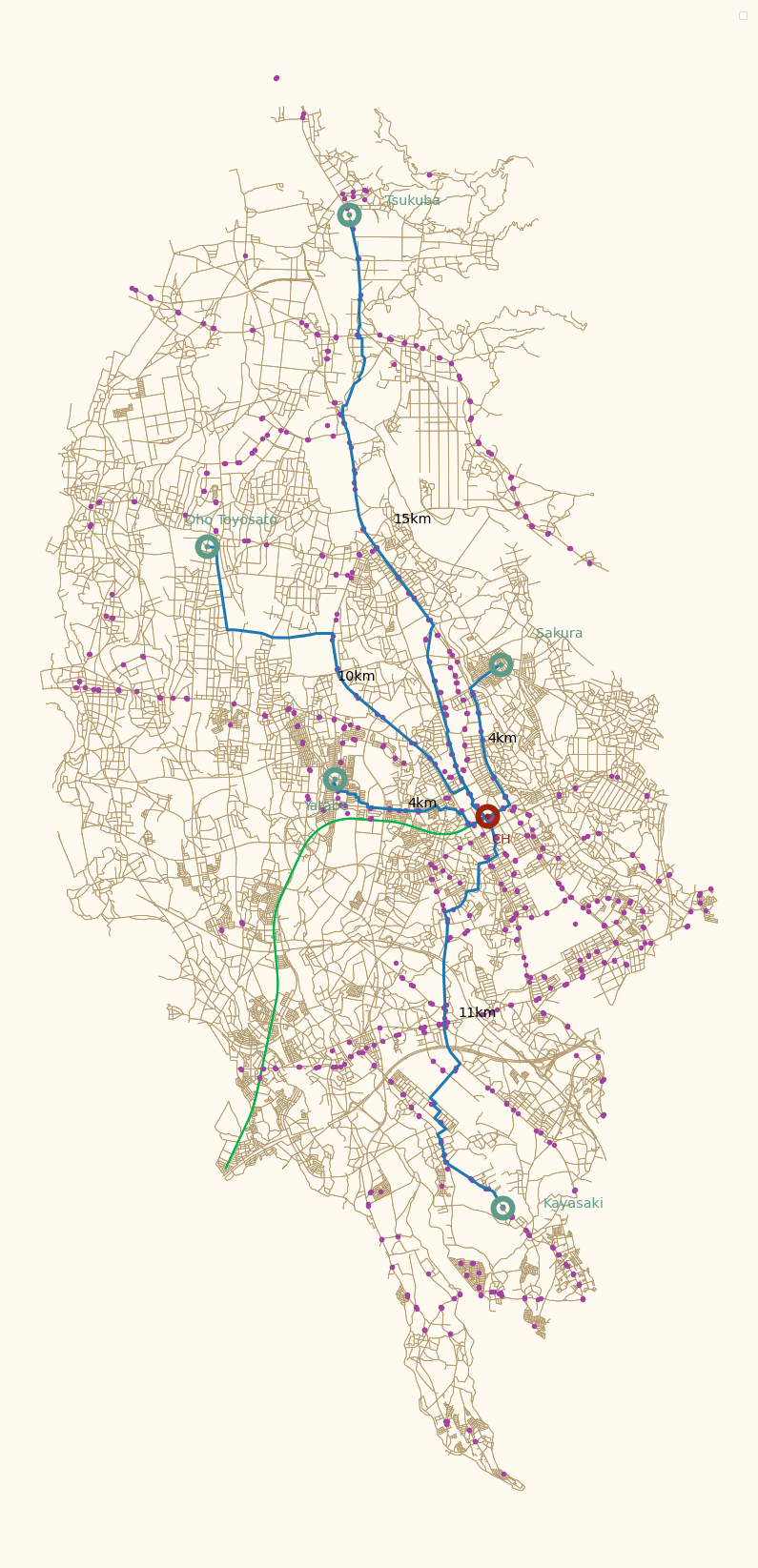}
    \caption{Map of the Tsukuba region. Tan lines are the drivable road system. The green line indicates the railway. The purple lines and dots represent the routes and stops of public buses respectively. The thick blue circles are each hub's PnR station. The thick red circle represents the central hub to which all other hubs are connected. The blue lines between each hub station and city center are the shortest route between them.}
    \label{fig:tsukubav2}
\end{figure}

In the Park-and-Ride system, we geographically divide Tsukuba into five districts and imagine the central placement of ``hubs'' in which customers park and from which they ride to a conveniently located hub with access to a variety of businesses and transportation to farther places (Figure \ref{fig:tsukubav2}). 

Using district maps and our dataset, we were able to subset the appropriate zone codes that make up the Tsukuba region. Also, using the Person Trip Survey dataset, we have an estimate of the number of trips departing from each hub in a day. These are:
\begin{enumerate}
\item[] \emph{Hub} $1 \rightarrow$ Tsukuba district identified by area code 5210, with 47,294 departures in a day,
\item[] \emph{Hub} $2 \rightarrow$ Oho Toyosato district  identified by area codes 5211 and 5212, with 93,269 departures in a day,
\item[] \emph{Hub} $3 \rightarrow$ Yatabe district identified by area codes 5213 and 5214, with 263,075 departures in a day
\item[] \emph{Hub} $4 \rightarrow$ Sakura district  identified by area codes 5215 and 5216, with 145,635 departures in a day, and 
\item[] \emph{Hub} $5 \rightarrow$ Kayasaki district  identified by area code 5217, with 39,843 departures in a day.
\end{enumerate}

As our study focuses on the public bus and private car user populations, we further subsetted the Tsukuba data to only include respondents (along with any companions) who used car and/or bus on the day of the survey.  Also, to make the five subsets more consistently sized and concise, trips were grouped into four-hour buckets by hour of departure. For each time bucket, we summed the trip expansion factors and divided the total by four so that it may represent the number of customers arriving at a PnR station per hour within a particular time bucket. The trip durations were averaged by all trips occurring withing the same hour.

From the Person Trip data, we have the current estimated average travel and wait times (in sum) for each response regarding trip duration in the survey. These are be assumed to be E$[T_{curr}]$, the average trip duration parameter needed to compute the vehicle density per kilometer $k_j$ (see Subsection \ref{subsec:approxkj}). 

For each customer arriving at a PnR station (or the central hub), we assign them to a single passenger private car with probability $p$ or to a public bus of capacity $C_{bus}$ with probability $1-p$. For our experiment, we simulate scenarios for $p$ ranging from 0 to 1, while we keep the bus capacity fixed at 100 individuals per bus (i.e. $C_{bus} = 100$). This capacity can be conveniently adjusted based on reasonable policies by public bus companies, such as overall change in bus capacity or setting different capacities based on region density. We provide insights on varying bus capacity and frequency by running our approximation model on various scenarios in Section \ref{sec:policyapp} while exploring data-driven transportation policies.

It should be noted that the number of customers being served, whether they are people arriving at the PnR or vehicles arriving at the service station, must stay below the stability condition allowance so that the simulation can terminate for a given set of parameters. As we wish to consider many possible parameter combinations, we programmatically restrict the number of customers and justify this restriction as ``give-ups'' - that is, a customer chooses to leave the PnR or the road if they cannot be served within an hour.

As the queuing model simulates the traffic between two stations via a single road, we set the distance $d_n$ between each pair of stations (from one district to the central hub) as the shortest-drivable-path length between them. These distances are as follows: 15 km between Tsukuba district and the central hub, 10 km between Oho Toyosato district and the central hub, 4 km between Yatabe district and the central hub, 4 km between Sakura district and the central hub, and finally 11 km between Kayasaki district and the central hub.  

We also need the nominal speed of each vehicle as an initial input to the queueing algorithm, $v_{NS}$. This number would ideally reflect the speed limits set in each road of the paths between the station pairs, however segmenting the paths would add extra complexity to the model. In order to keep the simulation simpler, we set the nominal speed according to a grid search to maximize the number of served customers (i.e. number of customers at or below the stability condition allowance). The grid ranges from 50 to 80 km/hour which we assume represents average of the speed limits from the roads in each path. Along with $k_j$, $v_{NS}$ is used to compute the road segment service rate $\mu$ in vehicles per hour.

The remaining parameter needed to run the queueing simulations, $b_n$ the inter-arrival time of buses at the $n$th PnR station in hours, is also obtained via a grid search ranging from 0.05 up to 0.5 so to maximize the number of served customers while considering the stability constraint. As explained earlier, the queueing model need to respect stability conditions (\ref{busstability}) and (\ref{roadstability}), which means that the bus inter-arrival time cannot be greater than 1 over the number of buses needed to fulfill the hourly demand at a particular PnR station. It is possible, and likely that when $p$ is closer to 1, a bus will arrive at the PnR when there are no bus customers. In that case, the bus will run empty as the $b_n$ parameter remains fixed after being set. This assumption helps us keep the model simple, but it also makes sense in reality, as bus schedules should remain predictable to their customers.

Once parameters are set, the simulation proceeds as described in \ref{MCsim}. Note that once the speed of a vehicle is computed, it serves as input into the MEET model described in Section \ref{sec:meet}. This is a valuable coupling of our model as it directly ties travel times with vehicle emissions. The simulated emissions functions, as in MEET, also takes as input: $d_n$ - the distance traveled from PnR to city center, vehicle type - decided by probability $p$ and bus capacity $C_{bus}$, and fuel type - a Bernoulli random variable where the probability of being a gasoline vehicle $p_{gasoline}$ is informed by estimated ratios from the \cite{veh2018owned}.

\subsection{Result analysis}\label{sec:results}

\begin{figure}
    \centering
    \includegraphics[width=\textwidth]{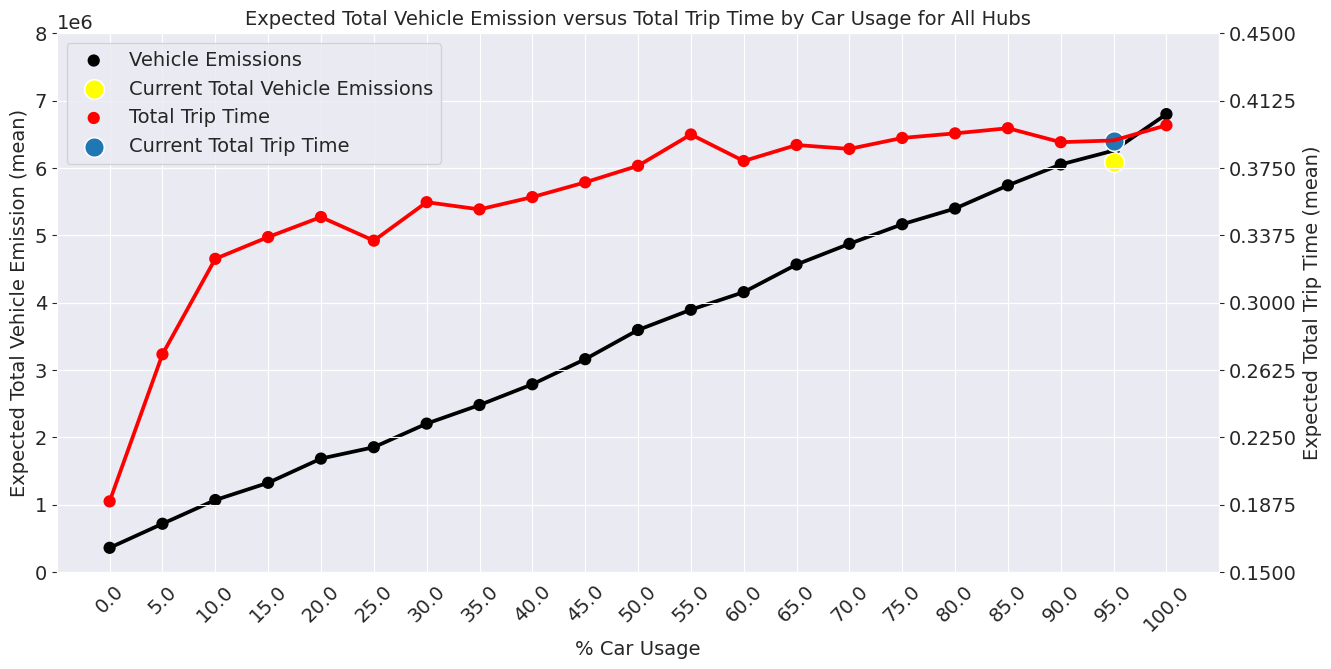}
    \caption{The Tsukuba PnR system. Plot shows expected vehicle emissions (back points and line) and total trip times (red points and line) for percentage of car usage. Current vehicle emissions and current mean trip time points are included for comparison (yellow and blue points respectively). Note the different scales for emissions (in grams of total pollutants) and time (in hours). Note that ``mean'' indicated on the axis is for aggregation over PnR hubs, direction of travel, and four hour time intervals, see Section \ref{numerical} for details on how the system was set up.}
    \label{fig:compwithcurr}
\end{figure}

Our PnR system and queue simulation yield some interesting results for the the case-study of Tsukuba. Figure \ref{fig:compwithcurr} show us the expected total vehicle emissions in grams of pollutants for the types of buses and cars considered in our simulation per percentage of car usage (red plot). In the same figure we can see the expected total trip time in hours for customers traveling between the PnR stations and the city center per percentage of car usage (black plot). As a test of simulation results, we also computed the empirical values for the ``current'' total emissions and expected travel times. We computed these values using the 2018 Person Trip survey dataset, subsetted to consider traffic within, to, and out of Tsukuba. As in the data used for the simulation, we also only considered customers using either private car or bus. Taking these types of customers as the population, we have that 95\% of customers used private cars instead of buses in 2018.

Our simulation does not allow customers to wait in the queues for more than one hour due to the corrections made so that the stability conditions are respected. For that reason, to compute the empirical expected total trip time, we only considered trips no longer than 60 minutes. Fortunately, for 90\% of the customer population trips took less than 60 minutes. The expected total trip time for the set of customers who use 95\% car/5\% bus and who take $\leq 60$ minutes trips was computed to be 0.4249 hours. This value falls a bit higher than what the simulation would expect, however this is not surprising since reality does not have such ideal PnR system and trips may take longer on average. We adjust for reality's lack of idealism by only considering trips no longer than 55 minutes, and plot the very close empirical expected total trip time of 0.3893 hours on Figure \ref{fig:compwithcurr} in blue.

Simplifying assumptions are necessary to compute an empirical estimate for emissions. Bus capacity is assumed to be 100 customers. To obtain an average number of buses per hour passing through the PnR system, from the aggregated Person Trip data for trips of this type, we assume that 20 customers share the same bus at a time. This leads to $b_n=0.0625$ for the bus inter-arrival time. For private cars, we assume an average of 1.25 customer per car per hub per time. From these assumptions and using data from the Person Trip survey, we can compute the total number of private cars and buses when private car usage is 95\%. We also assume an average speed of 70 km/hr for buses and 80 km/k for private cars, and take the average of all five PnR - City Center distances. Thus, given total number of vehicles, vehicle types, mean speeds, and mean distance, using MEET (Section \ref{sec:meet}) we obtain an estimate of 6,093,234 grams of pollutants emitted at 95\% car usage out of all bus and car customers in 2018. This estimate is plotted in yellow on Figure \ref{fig:compwithcurr}.

\begin{figure}[H]
    \centering
    \includegraphics[width=\textwidth]{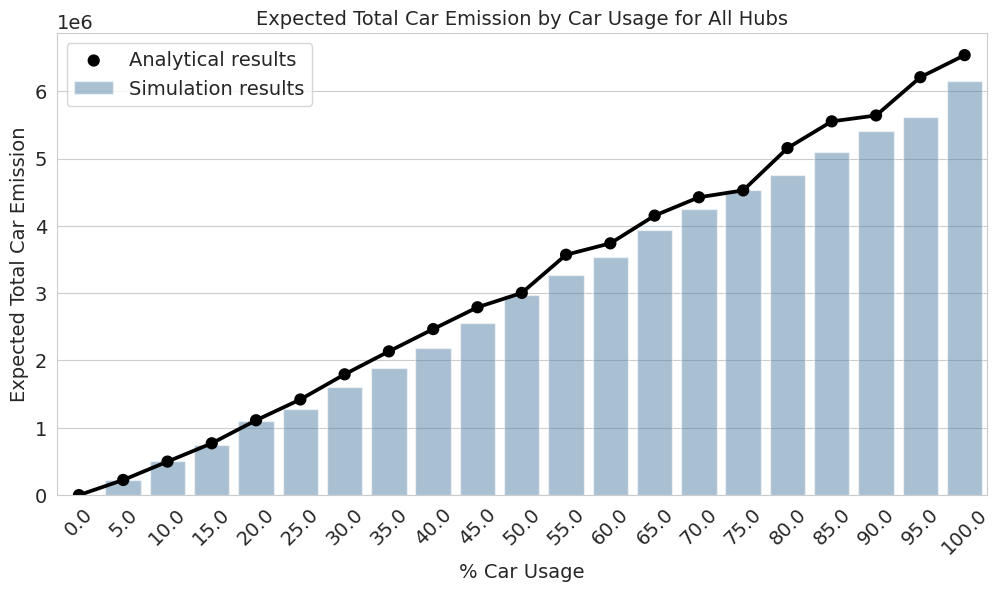}
    \caption{Total car emissions in grams averaged for all hubs using 4-hour intervals. Vertical bars represent simulation results while points and line represent the approximation model's analytical results. Analytical results for the arrival of cars to the service station use Erlang distribution with phase 20.}
    \label{fig:allhubsce}
\end{figure}

\begin{figure}[H]
    \centering
    \includegraphics[width=\textwidth]{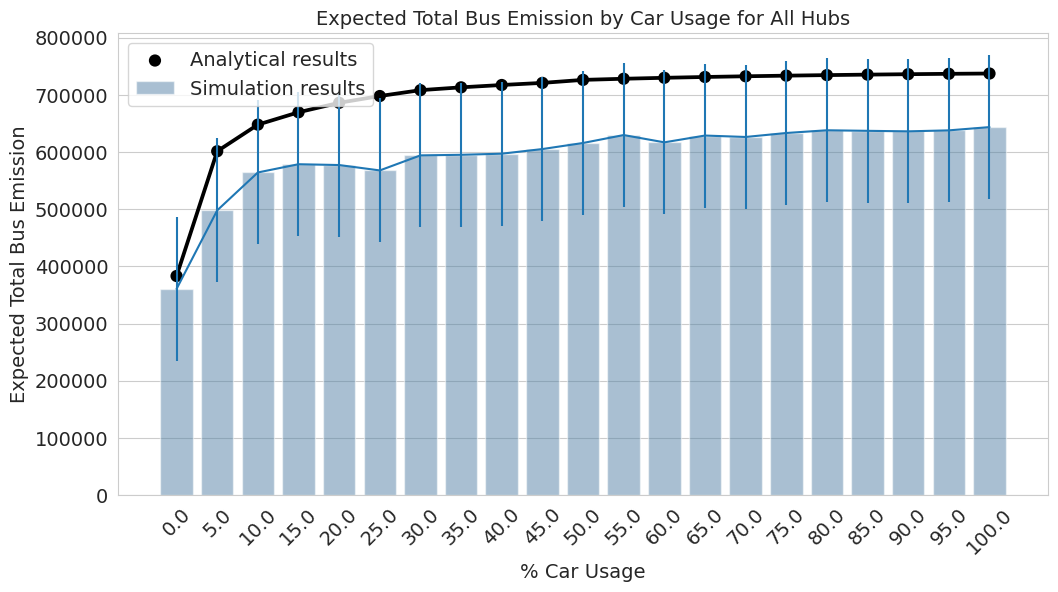}
    \caption{Total bus emissions in grams averaged for all hubs using 4-hour intervals. Vertical bars represent simulation results while points and line represent the approximation model's analytical results. Analytical results for the arrival of buses to the service station use Erlang distribution with phase 200. We have included 95\% confidence interval bars to demonstrate that analytical results fall inside the 95\% confidence interval of the simulated results.}
    \label{fig:allhubsbe}
\end{figure}

\begin{figure}[H]
    \centering
    \includegraphics[width=\textwidth]{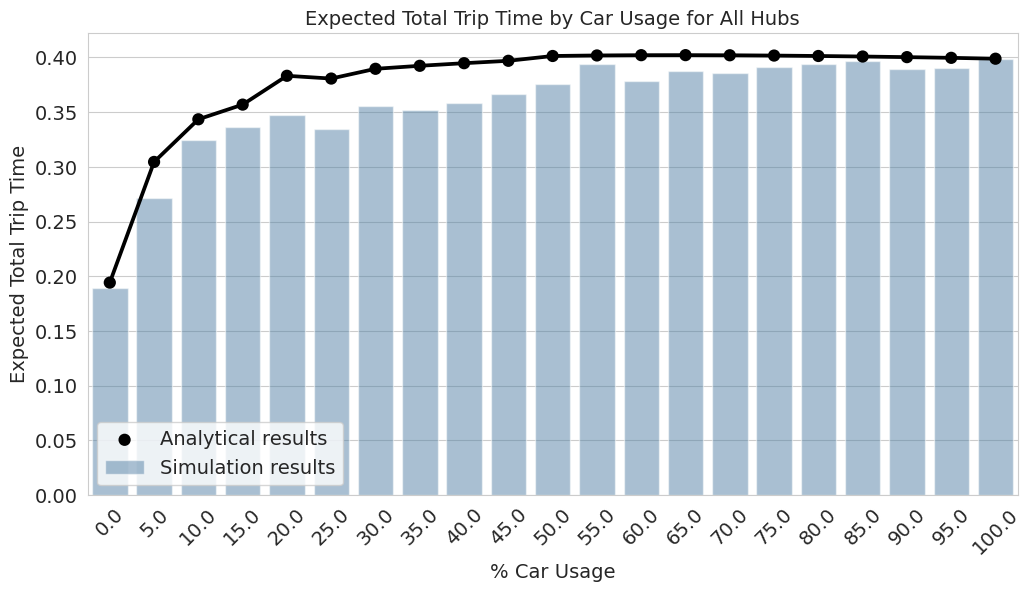}
    \caption{Total trip time in mean hours averaged across all hubs and over the 4-hour interval. Vertical bars represent simulation results while points and line represent the approximation model's analytical results.}
    \label{fig:allhubstt}
\end{figure}

\begin{figure}[H]
    \centering
    \includegraphics[width=\textwidth]{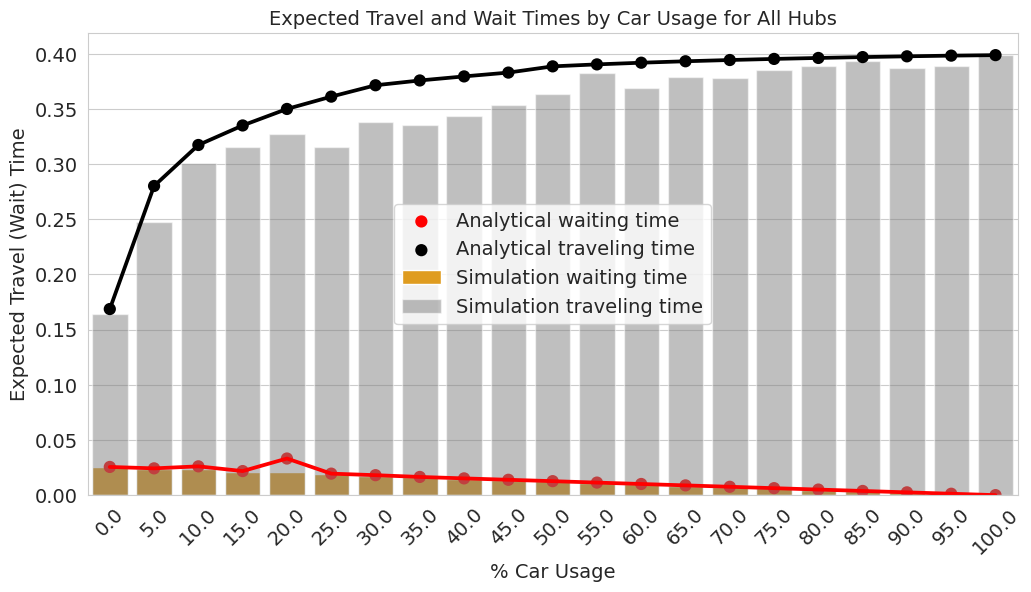}
    \caption{Traveling time and waiting time in mean hours averaged across all hubs and over the 4-hour interval. Vertical bars represent simulation results while points and line represent the approximation model's analytical results.}
    \label{fig:allhubswt}
\end{figure}

Results comparing our queue simulation and the approximation model using the Tsukuba case-study as input data are also obtained. Figures \ref{fig:allhubsce} and \ref{fig:allhubsbe} show agreement between results obtained from the simulation and approximation model for both private car and bus emissions using 4-hour interval averages, as do Figures \ref{fig:allhubstt} and \ref{fig:allhubswt} for the simulated and analytical mean over the 4-hour interval for expected total trip, traveling and waiting times. It should be noted that some discrepancy in Figure \ref{fig:allhubsbe} is attributed to the difference of the settings between the approximation model and the Monte Carlo Simulation, which is described in the first bullet point in Subsection \ref{subsec:setting} -- as an input of MEET, the actual sojourn time distribution in the service station is used in the simulation while the expected sojourn time is used in the approximation model. However, it should also be emphasized that almost all of the approximated values are within the 95\% confidence interval of the simulation experiment in Figure \ref{fig:allhubsbe}. Additionally, compared with the simulation results, the approximation model captures the safe side (i.e., slightly higher values) of the expected total bus emission. This tendency implies the validity of our proposed approximation model in addition to the benefit of reduced calculation cost.

Figure \ref{fig:allhubswt} overlays both traveling and waiting times to depict their trend relationship. Although the magnitude of the difference is high for this scenario -- since these results are from a PnR system with large and frequent buses, waiting times are very small compared to traveling times -- we can naturally observe that the expected waiting time becomes longer as the car usage rate becomes smaller, which provides compelling evidence in support of \ref{h1}. The opposite tendency can be confirmed for the traveling time and emissions, which shows support for \ref{h2}. In summary, it is evident that there exists a trade-off relationship between the waiting time and the traveling time (and accompanying emissions). 
If we introduce smaller and less frequent buses compared to the scenario in Figure \ref{fig:allhubswt}, it might be possible that the traveling time and emission can be reduced since the traffic congestion is alleviated and small capacity vehicles also discharge small emissions.
On the other hand, the risk in which the waiting time for customers explosively increases in such a scenario should not be dismissed at the same time. Therefore, it is clearly necessary to consider a framework that comprehensively evaluates all measures with a trade-off relationship (the waiting time, traveling time, and emissions) to determine the optimal transit policy, i.e., the optimal capacity and frequency of buses.
In the next section, we explore the discussion of the optimal transit policy for various car usage rates, using SCETT  -- the newly proposed total social cost in Subsection \ref{subsec:totalcost} that embraces all the above three measures.

\section{Transit policy application}\label{sec:policyapp}
\label{sec:transitpolicy}
Our PnR system combined with our SCETT model offers great potential to assist transit policy makers. We applied (\ref{eq:argminbC}) by using the Person Trip dataset as prepared for the Tsukuba case-study and the approximation model's results for trip times and carbon emissions, to obtain bus frequency intervals $b$ and bus capacity $C$ which minimize the social cost of emissions and trip time as per the FUND climate economy model: 
\begin{equation}
    \argmin\limits_{(b,C)}^{h,p} \left(\sum\limits_{d(h,p)} \sum\limits_{i(h,p)} \left( 8.2\times 10^{-6} \times E^{CO_2}_{(bus+car, h, 4)}(p, b, C) + 42.6 \times 4 \times \Delta T_{h}(p, b, C) \right) \right). \label{eq:tsukubCfund}
\end{equation}

\begin{figure}
    \centering
    \includegraphics[width=0.8\textwidth]{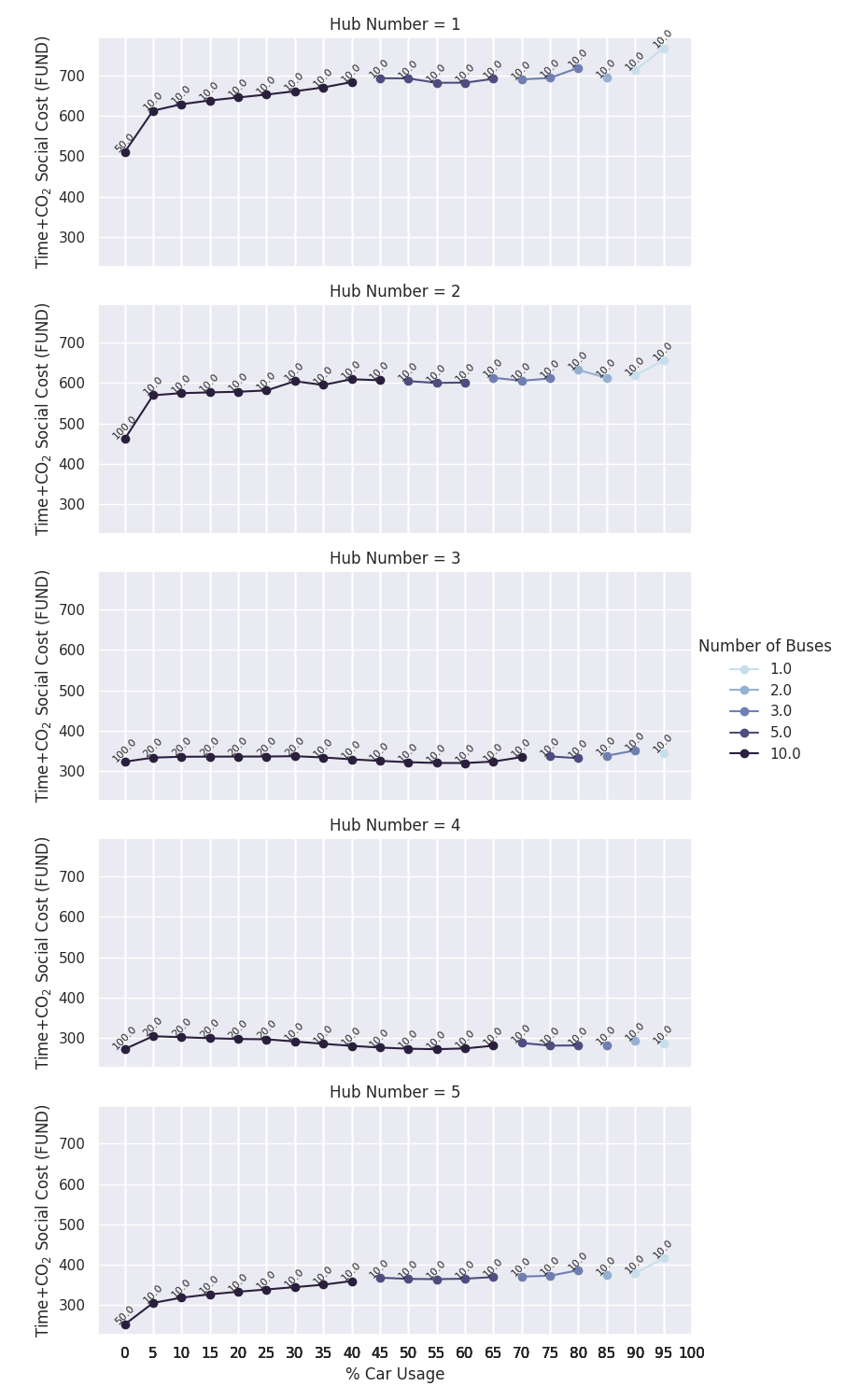}
    \caption{SCETT according to FUND in international dollars/capita per percent of car usage for a 24 hour time interval for each hub of the Tsukuba PnR system according to the 2018 Person Trip survey. Each point's color pertain to the number of buses ($[1/b]$) and annotations above each point represent the bus capacity. Both number of buses and bus capacity have been selected to minimize SCETT under the FUND model assumptions.}
    \label{fig:fund}
\end{figure}

In Figure \ref{fig:fund}, we visualize the results we obtained using (\ref{eq:tsukubCfund}) for each hub by holding the sub-region parameter fixed at a hub and then finding argument minima $b$ and $C$, where $b$ is in the linear interval containing 10 values from 0.1 to 1 and $C$ is in the set $\{10,20,30,...,100\}$. In each hub, for each car usage percentage ($p \times 100 \%$) we see what number of buses at what capacity should be deployed.

\begin{figure}
    \centering
    \includegraphics[width=0.8\textwidth]{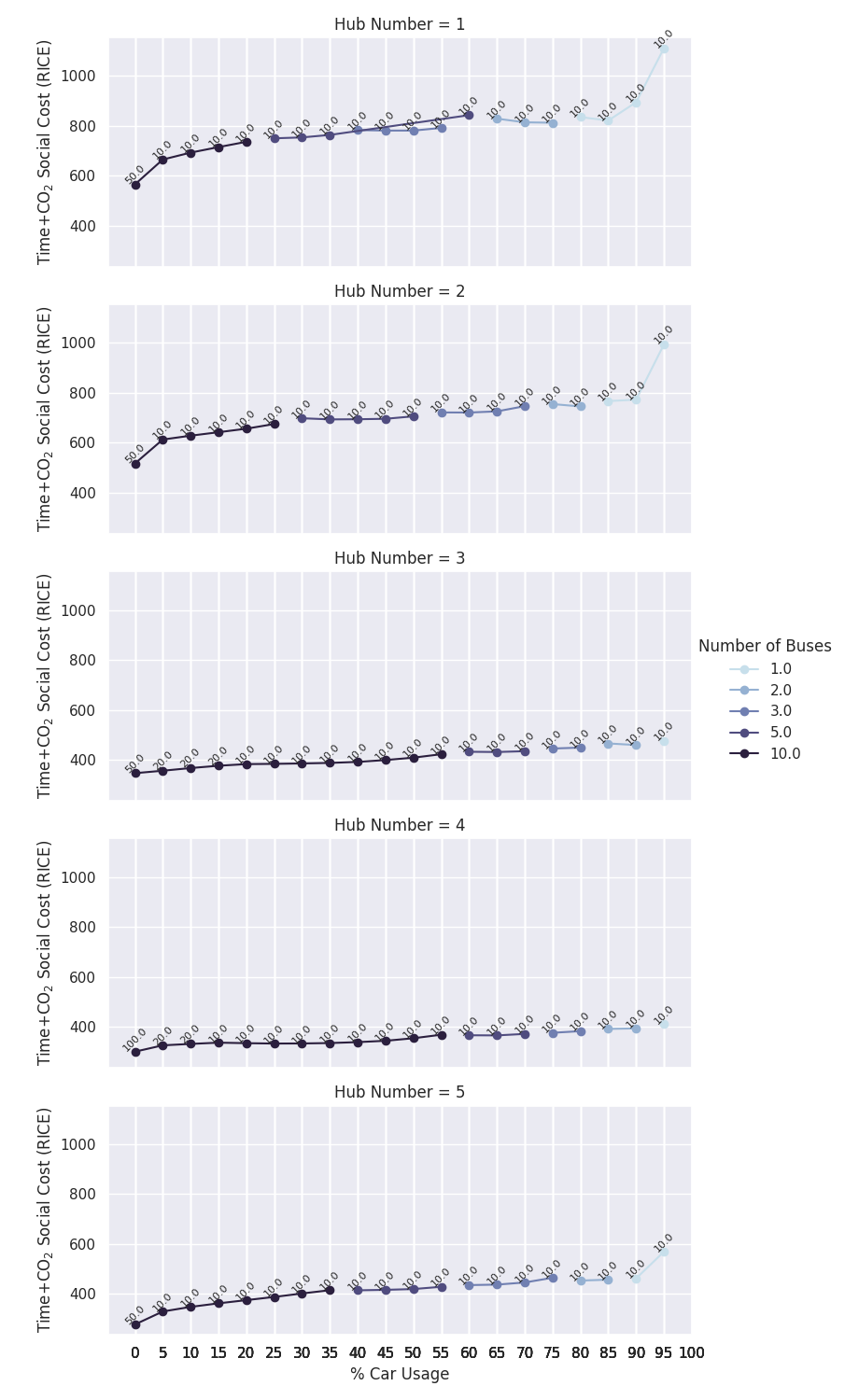}
    \caption{SCETT according to RICE in international dollars/capita per percent of car usage for a 24 hour time interval for each hub of the Tsukuba PnR system according to the 2018 Person Trip survey. Each point's color pertain to the number of buses ($[1/b]$) and annotations above each point represent the bus capacity. Both number of buses and bus capacity have been selected to minimize SCETT under the RICE model assumptions.}
    \label{fig:rice}
\end{figure}

Similarly for the RICE model we obtain:
\begin{equation}
    \argmin\limits_{(b,C)}^{h,p} \left(\sum\limits_{d(h,p)} \sum\limits_{i(h,p)} \left( 34.0\times 10^{-6} \times E^{CO_2}_{(bus+car, h, 4)}(p, b, C) + 42.6 \times 4 \times \Delta T_{h}(p, b, C) \right) \right), \label{eq:tsukubCrice}
\end{equation}
and in Figure \ref{fig:rice}, we visualize the results we obtained using (\ref{eq:tsukubCrice}).

As mentioned in Section \ref{subsec:totalcost}, the results represent daily bus frequency and capacity recommendation, despite the fact that for example 12PM and 12AM traffic can be quite different. Regardless of the time basis we use in our optimization model, we believe the coarse results we show are still a fair demonstration of the use and power the model. Also pragmatically, bus schedules may become too complex and hard to understand if both bus frequency and capacity change on an hourly or even 4-hourly basis. If we attempt to find solutions that match the dynamics of customer demand too finely, we risk losing model parsimony.

The cost trends appear similar for both the FUND and RICE integrated SCETT models, with mostly the magnitude of the cost showing their differences. We note that the trend is slightly different between models for hubs 3 and 4. The reason for this difference comes from the magnitude of the ratio between carbon cost and the trip time cost. The smaller the ratio, the greater the emphasis SCETT places in the trip time cost. Thus, we can see a trend difference in the solutions for hubs with an overall higher customer demand, such as hubs 3 and 4.

In both models, we note that smaller and more frequent buses provide more social benefit on average, with larger capacity vehicles being necessary only when the percentage of car use is low. This makes sense since when private car use is low, there is more demand for buses. In such scenario, if bus capacity is insufficient, then social cost due to travel time increases.

When private car use is high, buses can be small and infrequent. Since the demand for buses is low, the time cost of waiting for a bus is marginal. However, due to congestion caused by the high use of private cars, deploying more buses would only increase congestion and therefore travel time and vehicle emissions. Also, deploying larger buses would increase the baseline cost of emissions and not necessarily provide time cost reduction due to the low bus demand.

From the PnR perspective, we also observe interesting differences between the SCETT values in different PnR hubs. Whether we look at SCETT under FUND or RICE, we notice that for hubs 1 and 2 the baseline social cost is higher than for the remaining hubs. A primary reason for this difference is that PnR locations in hubs 1 and 2 are the farther from the city center, and for travel time and emissions computation, distance matters. This point raises the importance of how a PnR system set-up (e.g. where stations are placed) contributes to these social costs. A secondary reason is the populousness/demand of the PnR regions. In Section \ref{numerical}, we list the number of departures from each hub - listing the hubs in decreasing order of departures we have: 3, 4, 2, 1, and 5. Note that in hubs 3 and 4 the frequency of buses is not so promptly decreased as in the other hubs, and yet the social cost does not increase by much. This is due to the model balancing the emissions cost incurred from having more buses, with the savings in travel time cost of a large set of customers.

Furthermore, it is worth noting that SCETT takes relatively higher values in most hubs for the current car usage rate (95\%) compared to the lower car usage rates even under the optimal transit policies for each car usage rate. This tendency leads us to presume the limits of road handling capacity for the current high car usage rate. In particular, it can be observed from the graphs that the roads overflow, even for the optimal transit policies in hubs 1 and 2, where the number of departures is not as high as in 3, 4 or 5. Thus, our numerical experiments on SCETT can be regarded as evidence of the importance of having both optimal mass transit operation and a reduction of single-occupancy car use. The latter can be achieved via joint efforts between public and private sectors, through higher taxes on fossil fuels for private cars, lower prices for public transit, and addition of comfort and services for customers of public buses.

\begin{table}[h]
    \centering
    \caption{Current SCETTs per day per hub in international dollars for the FUND and RICE models. The values assume 90-95\% car users, and large buses (capacity $\geq 60$) deployed every 0.0625 hours.}
    \label{tab:curr_scett}
    \begin{tabular}{ccc}
    \hline
         Hub Number &  Time+CO$_2$ Social Cost (FUND) & Time+CO$_2$ Social Cost (RICE)\\
         \hline
         1& 1024.03& 1671.46\\
         2& 1053.50& 2239.84\\
         3& 1662.02& 4738.93\\
         4& 1011.24& 2708.66\\
         5& 553.23& 857.21\\
         \hline
    \end{tabular}
    
\end{table}

\begin{figure}
    \centering
    \includegraphics[width=\textwidth]{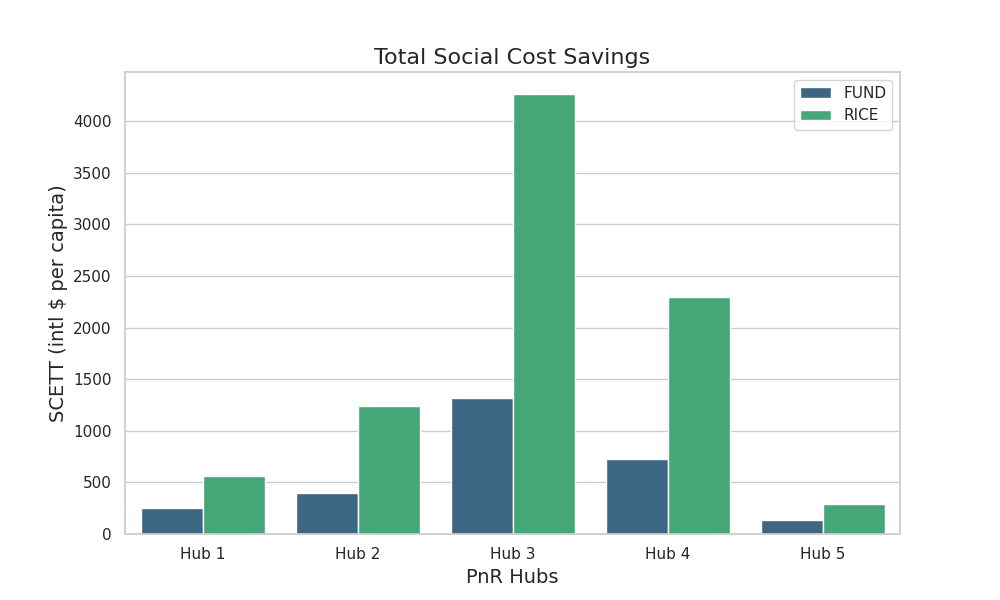}
    \caption{Total savings for the SCETT per day per hub under FUND and RICE models' assumptions. This saving are from having a PnR system as described in this study and bus policies that match the optimum SCETT for 95\% car use percentage.}
    \label{fig:SCETTsave}
\end{figure}
For comparison, we estimated the current SCETT for each hub by obtaining the proportion of private car users from the Person Trip survey dataset (recall that our customer population use either private cars or buses), and assuming buses of capacity 60-100 arriving every 0.0625 hours (see Table \ref{tab:curr_scett}). Looking at Figures \ref{fig:fund} and \ref{fig:rice}, it is clear that we can have significant social cost savings by making data-driven decisions on transit policies. Notably, these findings directly support \ref{h3}, which proposed that PnR system optimization involves multi-dimensional trade-offs between travel times, waiting times, vehicle emissions, and social costs. Total savings for the SCETT under FUND and RICE models' assumptions are depicted in Figure \ref{fig:SCETTsave}. These savings were computed assuming the PnR system described in this study and bus policies that match the optimum SCETT for 95\% car use percentage (current private car usage). Savings can be more significant if there are less private car users in the PnR system, but they were kept at current level car usage for comparison purposes.

Thus far, we have proposed a mathematical method to derive the optimal frequency and capacity of the bus in PnR model. The results empirically validate the complex trade-off relationships predicted by \ref{h3}, demonstrating how adjusting system parameters simultaneously impacts multiple performance dimensions. There is a risk that inappropriate setting of the bus leads to lower social welfare, i.e., long waiting and traveling time and high emissions. However, we have shown that our proposed methodology drastically reduces the social cost through the case studies. Undoubtedly, these results and our model in general, could provide further insights if we were to look at the demographics of these hubs, consider other types of vehicles, or experiment with varying PnR set-ups. Using our model ensemble would allow governments, automobile makers and bus companies to come together in the creation of transit policies of benefit to society by setting parameters that keep GDP high and pollutant emissions low.


\section{Conclusion}
\label{sec:conclusion}

In this study, we presented a comprehensive mathematical model for evaluating the performance measures of traveling and waiting times, as well as vehicle emissions, within the context of a Park and Ride (PnR) system. Our model facilitates the calculation of these performance metrics across various scenarios, considering the usage rate of the PnR system. We examined scenarios where individuals opt to travel in their personal vehicles without wait or in public buses, which may involve waiting times, and where vehicles navigate congested or clear road conditions. By simulating a range of scenarios with varying percentages of single occupancy personal vehicle usage, we investigated the relationship between total trip time for customers (sum of waiting and traveling times) and vehicle emission levels. 

In particular, we showed the positive correlation between traveling times and emissions, a result which supports hypothesis \ref{h2}. We verified the negative correlation between waiting times and traveling times (and accompanying emissions) as predicted by \ref{h1}. That is, there exists a trade-off relationship between the waiting time and the traveling time -- the traveling time (and emissions) can be reduced by introducing small capacity and infrequent public buses as traffic congestion is alleviated while the waiting time is increased, and vice versa for large and frequent public buses, this result strongly supports \ref{h3}. Considering this trade-off tendency, the primary objective of this paper was to explore transportation policies aimed at minimizing the total time and emission social costs within a particular region, with a focus on the implications of PnR system considerations. This study fills a significant gap in the literature by integrating all the interdependent elements – time and emission costs – within a unified framework, offering valuable insights for transit infrastructure administrators contemplating PnR system implementation.

To achieve our objectives, we proposed a novel approach that combines two distinct queues – one representing the movement of people and the other vehicles – building upon existing literature on queueing theory. Our methodology integrated Monte Carlo simulation techniques to compute travel and wait times for customers, while emissions computations leveraged established methodologies. The validity of our approach was verified through a comparison between simulation and analytical results, ensuring the robustness of our findings. We also introduced a modified total social cost function that allows for separate parameters to account for social emissions costs and economic productivity costs (which depend on the time cost) for different percentages of private car usage, and different capacity and frequency of public buses. As a highlight of this study, we further proposed the method to determine the optimal transit policy, that is, the optimal capacity and frequency of buses for various single-occupancy car usage rates. It is promising that the proposed optimization method on transit policy allows governments, automobile makers and bus companies to come together in the creation of a transit scheme which benefit society, by setting parameters that keep GDP high and pollutant emissions low.

We applied our methodology to a case-study focusing on Tsukuba city in Japan, which provided insights into the practical implications of our approach. By analyzing empirical data from the Person Trip Survey conducted in the Tokyo metropolitan area, we refined our model parameters and validated our simulation results. We also offered actionable policy recommendations by computing the total social cost incorporating travel time and CO$_{2}$ emissions, facilitating informed decision-making for transit policies. With the development of autonomous vehicles and advanced vehicle control technology, it is promising that the flexibility of the transportation system, such as frequency and capacity, will be enhanced. Our proposed method to determine the optimal transit policy has potential to play an important role enabling us to lower both the trip time of customers and environmental damage for society.

Our numerical results under the optimal transit policies showed significant reduction in the total social cost for the current high car usage rate of 95\% in Tsukuba. Specifically, implementing optimal bus frequency and capacity policies could reduce total social costs by up to 30\% on average per hub, compared to current conditions (Figures \ref{fig:fund} and \ref{fig:rice}). Furthermore, the results indicate that even further reduction in the total social cost can be achieved by decreasing the single-occupancy car usage rate from the current state. For instance, at a 70\% car usage rate, the total social cost could be reduced by approximately 45-50\% under both the FUND and RICE models compared to the current 95\% rate (Figures \ref{fig:fund} and \ref{fig:rice}). This implies the social importance of promoting car-free movement and mass transportation. Overall, our research contributes to advancing the understanding of the complex dynamics between transportation systems, emissions, and societal welfare, paving the way for more sustainable and efficient urban mobility strategies.

Despite the contributions of this study, there are some limitations that should be acknowledged and addressed in future research. First, while our case-study focused on Tsukuba city, comparing the results with other cities could provide additional insights and validate the generalizability of our findings. Second, explicitly incorporating autonomous vehicles and electric vehicles into the model would enable a more comprehensive analysis of emerging transportation technologies. Third, considering waiting time and traveling time-sensitive demand for public buses by customers, that is, ``strategic'' customer behaviors such that proposed in \cite{hassin2003queue}, could further refine the model and capture more realistic behavior. Finally, while we highlighted the social optimization of PnR system in this paper, the revenue maximization of bus companies under the consideration of operating costs is also a crucial issue.

Another interesting direction for future research would be to adapt the current model to support demand-responsive transportation. Such adaptation could provide insights into how dynamic, demand-responsive transit services impact waiting times, travel times, and emissions within a PnR context. By modeling on-demand bus arrivals as a stochastic process dependent on customer demand rather than fixed intervals, researchers could explore the potential benefits and challenges of implementing such systems. This extension would require modifying the queueing model to account for the complex interactions between customer demand, bus dispatch policies, and road traffic conditions. Investigating PnR systems with on-demand buses could offer valuable insights for transit agencies seeking to optimize their services and improve the overall efficiency and sustainability of urban mobility.

In conclusion, this study presents a novel framework for evaluating and optimizing PnR systems, considering the intricate relationships between travel times, waiting times, and vehicle emissions. The proposed methodology and case study demonstrate the potential for data-driven decision-making in urban transportation planning, contributing to the development of more sustainable and efficient mobility solutions. Future research should build upon these findings, addressing the limitations and incorporating emerging technologies to further advance our understanding of urban transportation systems and their societal impacts.

\section*{Acknowledgement}
The authors appreciate the three anonymous reviewers for their insightful comments and constructive suggestions which improved the presentation of this paper. This study is supported by F-MIRAI: R\&D Center for Frontiers of MIRAI in Policy \& Technology, the University of Tsukuba and Toyota Motor Corporation collaborative R\&D center. In addition, this work is supported by Interdisciplinary Top-Level Human Resource Development Project on Mathematical, Data Science, and AI, University of Tsukuba. The first author is funded by JSPS KAKENHI Grant Number JP 23KJ0249. Special thanks go to Amanda E. Hampton for her facilitation and discussion during Graduate-level Research in Industrial Projects for Students (g-RIPS-Sendai 2022).

\appendix


\section{Reference tables and figures}

\begin{table}[H]
 \caption{Speed dependency of various emission factors for gasoline passenger cars of EURO I vehicle class with $1.4 < CC < 2.01$ cylinder capacity and speed range of 10-130 km/hr}
 \label{table:gaspc}
 \centering
  \begin{tabular}{ll}
   \hline
   Pollutant & Emission factor (g/km) \\
   \hline \hline
   CO  &  $9.617 - 0.245v + 0.001729v^2$ \\
   CO$_2$  & $231 - 3.62v + 0.0263v^2 + 2526/v$ \\
   VOC   & $0.4494 - 0.00888v + 5.21\times10^{-5}v^2$\\
   NO$_X$   &  $0.526 - 0.0085v + 8.54\times10^{-5}v^2$ \\
   PM       & $0.0$ \\
   \hline
  \end{tabular}
\end{table}

\begin{table}[H]
 \caption{Speed dependency of various emission factors for diesel passenger cars weighing less than 2.5 metric tons and having speed range of 10-130 km/hr}
 \label{table:dieselpc}
 \centering
  \begin{tabular}{ll}
   \hline
   Pollutant & Emission factor (g/km) \\
   \hline \hline
   CO  &  $1.4497 - 0.03385v + 2.1\times10^{-4}v^2$ \\
   CO$_2$  & $286 - 4.07v + 0.0271v^2$ \\
   VOC   & $0.1978 - 0.003925v + 2.24\times10^{-5}v^2$\\
   NO$_X$   &  $1.4335 - 0.026v + 1.785\times10^{-4}v^2$ \\
   PM       & $0.1804 - 0.004415v + 3.33\times10^{-5}v^2$ \\
   \hline
  \end{tabular}
\end{table}

\begin{table}[H]
 \caption{The coefficients of emission functions for heavy goods vehicles with gross vehicle weights from 3.5 to 7.5 metric tons in MEET}
 \label{table:caplt30}
 \centering
  \begin{tabular}{lccccccc}
   \hline
   Weight class & $K$ & $a$& $b$ & $c$ & $d$ & $e$ & $f$ \\
   \hline \hline
   CO  & 1.50 & $-0.0595$ & 0.00119 & $-6.16\times10^{-6}$ & 58.8 & 0 & 0 \\
   CO$_2$  &110 &0& 0 & 0.000375 & 8702 & 0 &  0 \\
   VOC   &0.186 & 0 & 0 & $-2.97\times10^{-7}$ & 61.5 & 0 & 0 \\
   NO$_X$   &0.508 & 0 & 0 & $3.87\times10^{-6}$ & 92.5 & $-77.3$ & 0 \\
   PM       &0.0506 & 0 & 0 & $1.22\times10^{-7}$ & 12.5 & 0 & $-21.1$ \\
   \hline
  \end{tabular}
\end{table}

\begin{table}[H]
 \caption{The coefficients of emission functions for heavy goods vehicles with gross vehicle weights from 7.5 to 16 metric tons in MEET}
 \label{table:caplt60}
 \centering
  \begin{tabular}{lccccccc}
   \hline
   Weight class & $K$ & $a$& $b$ & $c$ & $d$ & $e$ & $f$ \\
   \hline \hline
   CO  & 3.08 & $-0.0135$ & 0 & 0 & $-37.7$ & 1560 & $-5736$ \\
   CO$_2$  &871 &$-16.0$& 0.143 & 0 & 0 & 32031 &  0 \\
   VOC   &1.37 & 0 & $-8.10\times 10^{-5}$ & 0 & 0 & 870 & $-3282$ \\
   NO$_X$   &2.59 & 0 & $-0.000665$ & $8.56\times10^{-6}$ & 140 & 0 & 0 \\
   PM       &0.0541 & 0.00151 & 0 & 0 & 17.1 & 0 & 0 \\
   \hline
  \end{tabular}
\end{table}

\begin{table}[H]
 \caption{The coefficients of emission functions for urban buses in MEET}
 \label{table:capub}
 \centering
  \begin{tabular}{lccccccc}
   \hline
   Weight class & $K$ & $a$& $b$ & $c$ & $d$ & $e$ & $f$ \\
   \hline \hline
   CO  & 1.64 & 0 & 0 & 0 & 132 & 0 & 0 \\
   CO$_2$  &679&0& 0 & $-0.00268$ & 9635 & 0 &  0 \\
   VOC   &0.0778 & 0 & 0 & 0 & 41.2 & 0 & 184 \\
   NO$_X$   &16.3 & $-0.173$ & 0 & 0 & 111 & 0 & 0 \\
   PM       &0.0694 & 0 & 0.000366 & $8.71\times10^{-6}$ & 13.9 & 0 & 0 \\
   \hline
  \end{tabular}
\end{table}

\begin{table}[H]
\caption{\cite{anthoff2019inequality}'s SCC estimates when incorporating regional income disparities in both the FUND and RICE climate economic models ($\gamma = 0.7$). All estimates are in international dollar per metric ton of CO$_2$}
    \label{tab:scc}
    \centering
    \begin{tabular}{lcc}
    \hline
    Region& FUND & RICE \\
    \hline \hline
        Africa &0.5 &5 \\
        China & 1& 12\\
        EU & 6.75& 32\\
        Japan &8.2&34 \\
        Middle East & 1.2& 15\\
        South America &1.8 & 15\\
        USA & 7.6& 41.2\\
        \hline
    \end{tabular}
\end{table}

\begin{figure}
    \centering
    \includegraphics[width=\textwidth]{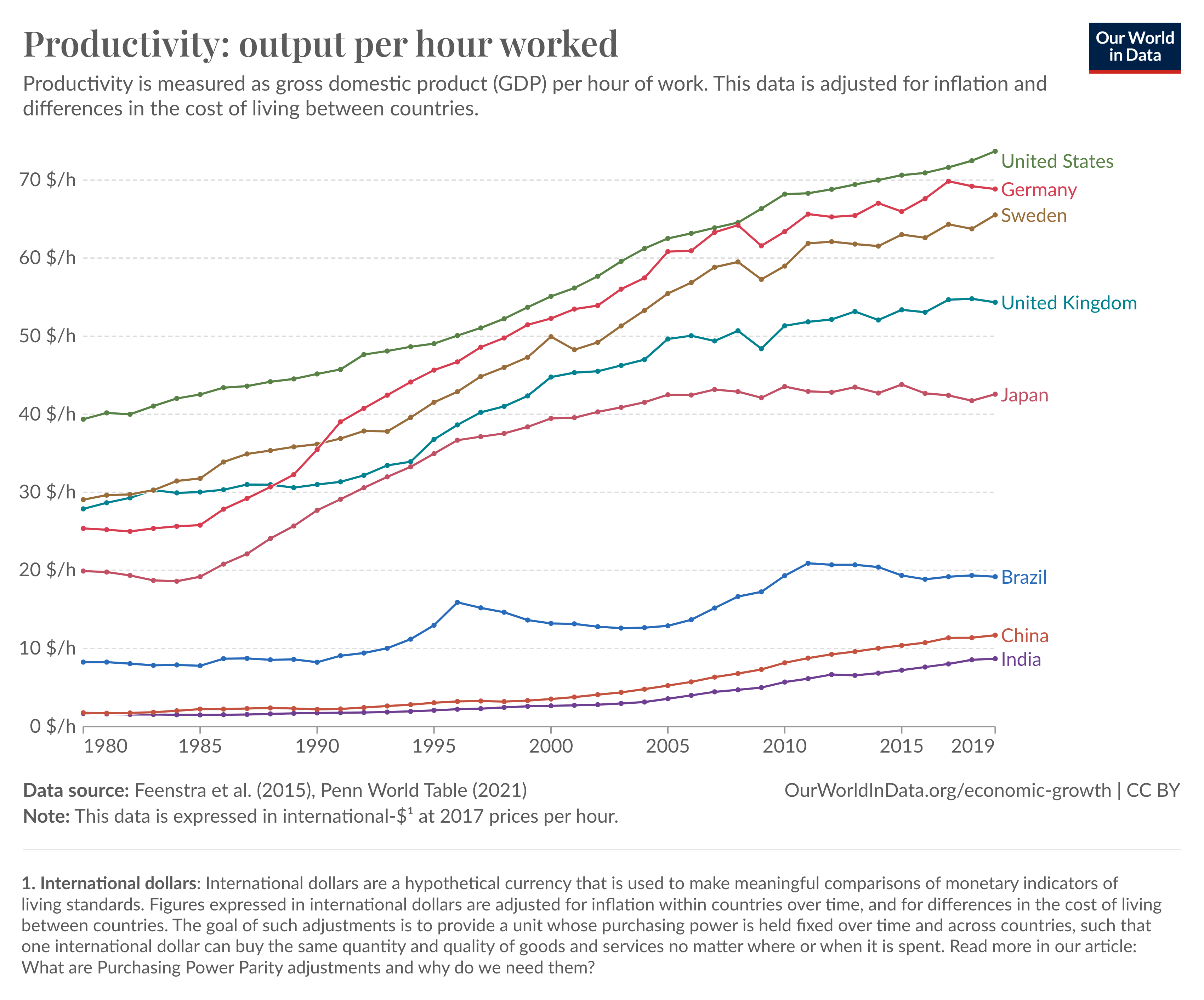}
    \caption{Trends in productivity measured in international dollars per hour from 1980-2019 for eight countries (\cite{feenstra2015next} with major processing by \cite{ourworld2024data})}
    \label{fig:prod}
\end{figure}

\begin{table}[ht!]
\caption{Pertinent categorical data collected from the Person Trip Survey of the Tokyo metropolitan area. For each member of the household, the dataset will also include features such as trip duration, number of trips on the same day, and number of companions. \label{pt_data}}
\centering
\begin{tabular}{l p{.6in} p{3in}}
\hline
Feature & Values & Description\\
\hline \hline
Zone &4-6 digits & This code indicates from high to low levels where a person is from, where they left from and where they are going to\\
\hline
Gender & 1-Male/ 2-Female& Sex at birth \\
\hline
Age group &17 buckets of 4 years& The age of household member who traveled starting from 5-85+\\
\hline
Occupation & 3 types & Primary industry, secondary industry, or other\\
\hline
Employment cat. & 9 types & Employment type ranging from self-employment to unemployed. Includes housewife and student. \\
\hline
Driver's lic. & 3 types &  Have, do not have, or returned  \\
\hline
Have car?&  3 types  & Own car, family car, or no car\\
\hline
Physical difficulties &types 1-5 & Not difficult at 1 including traveling with infant up to 5 basically unable to leave home\\
\hline
Household income &types 1-5 & Buckets from less to 2 million yen to more than 15 million yen\\
\hline
Facilities &15 types& The type of facility from which the person departed or left. Examples: school. park, commerce, office, etc. \\
\hline
Hours &0-24 o'clock& Times of arrival and departure from location to location, as well as shift start times for those whose trip is work related\\
\hline
Parking area & 6 types& Parking meter, long to short term parking, facility had parking, or no parking\\
\hline
Bicycle parking & 6 types& Parking on road, long to short term parking, facility had parking, or no parking\\
\hline
Transport type & 14 types & Rail, street car, local bus, express bus, personal car, trucks, car rental, private bus, taxi, bike, bike rental, private bike, walking, or other \\
\hline
Purpose type & 7 types & Activities ranging from going shopping to going to work or school\\
\hline
\end{tabular}
\end{table}
\newpage
\section{Analysis of QBD process for the service station on the road}
\label{QBD}
First, we define $\mathcal{L}_k$ $(k=0,1,2,\dots)$ as follows:
\[
\mathcal{L}_k = \{(i,q,r);i=k,\   q\in \left\{0,1,\dots,l_{q}-1\right\},\  r\in\left\{ 0,1,\dots,l_{r}-1\right\}\}.
\]
Then, the infinitesimal generator of QBD process, $\mathbf{Q}^{(L)}$, is given as follows:

\[\mathbf{Q}^{(L)}= \bordermatrix{
& \mathcal{L}_0 &\mathcal{L}_1 & \mathcal{L}_2 & \mathcal{L}_3 & \cdots	  \cr
\mathcal{L}_0 & \mathbf{B}^{(L)}_0& \mathbf{A}^{(L)}_1 & \mathbf{O}& \mathbf{O} & \cdots	 \cr
\mathcal{L}_1 & \mathbf{A}^{(L)}_{-1}& \mathbf{A}^{(L)}_0& \mathbf{A}^{(L)}_1& \mathbf{O} & \cdots	  \cr
\mathcal{L}_2 & \mathbf{O} & \mathbf{A}^{(L)}_{-1} & \mathbf{A}^{(L)}_0& \mathbf{A}^{(L)}_1 & \cdots	 \cr
\mathcal{L}_3& \mathbf{O} & \mathbf{O} & \mathbf{A}^{(L)}_{-1} & \mathbf{A}^{(L)}_{0} & \cdots	  \cr
\vdots & \vdots  & \vdots  & \vdots & \vdots  &  \ddots \cr
},\]
where

$\mathbf{A}^{(L)}_{0}=$
\scalebox{0.82}{
$\begin{pmatrix}
 A^{(L)}_{0, (0, 0), (0, 0)} & A^{(L)}_{0, (0, 0), (0, 1)}  & A^{(L)}_{0, (0, 0), (0, 2)} & \ddots & A^{(L)}_{0, (0, 0), (l_{q}-1, l_{r}-1)}   \\
 A^{(L)}_{0, (0, 1), (0, 0)} & A^{(L)}_{0, (0, 1), (0, 1)}  & A^{(L)}_{0, (0, 1), (0, 2)} & \ddots & A^{(L)}_{0, (0, 1), (l_{q}-1, l_{r}-1)}   \\
 A^{(L)}_{0, (0, 2), (0, 0)}  & A^{(L)}_{0, (0, 2), (0, 1)}  & A^{(L)}_{0, (0, 2), (0, 2)} & \ddots & A^{(L)}_{0, (0, 2), (l_{q}-1, l_{r}-1)}   \\
 \vdots & \ddots & \ddots& \ddots & \vdots  \\
 A^{(L)}_{0, (l_{q}-1, l_{r}-1), (0, 0)}  & A^{(L)}_{0, (l_{q}-1, l_{r}-1), (0,1)}  & A^{(L)}_{0, (l_{q}-1, l_{r}-1), (0, 2)} & \ddots & A^{(L)}_{0, (l_{q}-1, l_{r}-1), (l_{q}-1, l_{r}-1)}   \\
 \end{pmatrix}
$},\\ 
\\ 
and the same is true for the other matrices (i.e., $\mathbf{B}^{(L)}_0$, $\mathbf{A}^{(L)}_1$, and $\mathbf{A}^{(L)}_{-1}$), 
and $\mathbf{O}$ is a zero block matrix with an appropriate size. 
The elements of each matrix are written as follows (note that we have to define the diagonal elements properly so that the row sum of the infinitesimal generator $\mathbf{Q}^{(L)}$ equals to 0):
\[
A^{(L)}_{1, (q, r), (q, r)}  =  p \lambda_n, \qquad (q\in \left\{0,1,\dots,l_{q}-1\right\},\  r\in\left\{ 0,1,\dots,l_{r}-1\right\}),
\]
\[
A^{(L)}_{1, (q, l_{r}-1), (q, 0)}  =  \dfrac{l_{r}}{b_n}, \qquad (q\in \left\{0,1,\dots,l_{q}-1\right\}),
\]
\[
A^{(L)}_{-1, (l_{q}-1, r), (0, r)}  =  l_{q} \mu, \qquad (r\in \left\{0,1,\dots,l_{r}-1\right\}),
\]
\[
A^{(L)}_{0, (q, r), (q+1, r)}  = l_{q} \mu, \qquad (q\in \left\{0,1,\dots,l_{q}-2\right\},\  r\in\left\{ 0,1,\dots,l_{r}-1\right\}),
\]
\[
A^{(L)}_{0, (q, r), (q, r+1)}  =  \dfrac{l_{r}}{b_n}, \qquad (q\in \left\{0,1,\dots,l_{q}-1\right\},\  r\in\left\{ 0,1,\dots,l_{r}-2\right\}),
\]
\[A^{(L)}_{0, (q, r), (q, r)} =-\left(p \lambda_n +\dfrac{l_{r}}{b_n} + l_{q} \mu  \right),  \qquad (q\in \left\{0,1,\dots,l_{q}-1\right\},\  r\in\left\{ 0,1,\dots,l_{r}-1\right\}),  
\]
\[
B^{(L)}_{0, (q, r), (q, r+1)}  =  \dfrac{l_{r}}{b_n}, \qquad (q\in \left\{0,1,\dots,l_{q}-1\right\},\  r\in\left\{ 0,1,\dots,l_{r}-2\right\}),
\]
\[B^{(L)}_{0, (q, r), (q, r)} =-\left(p \lambda_n +\dfrac{l_{r}}{b_n}   \right),  \qquad (q\in \left\{0,1,\dots,l_{q}-1\right\},\  r\in\left\{ 0,1,\dots,l_{r}-1\right\}).
\]
Here, the elements which are not defined above are 0.

Based on the above, we derive the stability condition. We define the infinitesimal generator $\mathbf{A}^{(L)}$ of the phase as follows. 

\begin{equation*}
    \begin{split}
        \mathbf{A}^{(L)}&= \mathbf{A}^{(L)}_{-1}+ \mathbf{A}^{(L)}_{0}+\mathbf{A}^{(L)}_{1}.\\
    \end{split}
\end{equation*}
Assuming that the steady state probability of the phase is
\begin{equation*}
    \begin{split}
        &\mathbf{\eta}=(\eta_{(0,0)},\eta_{(0,1)},\eta_{(0,2)}, \dots, \eta_{(l_q-1,l_r-1)}),\\ 
    \end{split}
\end{equation*}
the following set of equation holds:
\begin{equation}
\label{tauA}
\mathbf{\eta} \mathbf{A}^{(L)}=\mathbf{0},
\end{equation}
\begin{equation}
\label{taue}
\mathbf{ \eta} \mathbf{e}=1,
\end{equation}
where $\mathbf{0}$ is a vector of zeros and $\mathbf{e}$ is a vertical vector of ones with an appropriate size, respectively.
Solving (\ref{tauA}) and (\ref{taue}), we obtain
\[
\mathbf{\eta}=\dfrac{1}{l_q l_r}\mathbf{e}^\top,
\]
where $\mathbf{e}^\top$ means the transpose of $\mathbf{e}$ (it is natural since the phase consists of the progressions of  two independent Erlang distributions with shapes $l_q$ and $l_r$).

Using $\mathbf{\eta}$, 
according to e.g., \cite{adan2017analysis}, the stability condition of QBD is given as follows:
\begin{gather*}
\mathbf{\eta} \mathbf{A}^{(L)}_{1}\mathbf{e} -\mathbf{\eta} \mathbf{A}^{(L)}_{-1}\mathbf{e}  < 0 \notag \\
\Longrightarrow \lambda_{n}  p +  \dfrac{1}{b_{n}}<  \mu,
\end{gather*}
which is same with the stability condition of the original model (\ref{roadstability}).

The steady state probability $\xi_{i,q,r}^\ast $ exists under the stability condition since our Markov chains are irreducible. We further define $\mathbf{\xi}_i^\ast $ and $\mathbf{\xi}^\ast $ as follows:
\begin{equation*}
    \begin{split}
        & \qquad  \qquad \mathbf{\xi}_i^\ast  = (\xi_{i,(0,0)}^\ast,\xi_{i,(0,1)}^\ast,\xi_{i,(0,2)}^\ast, \dots, \xi_{i,(l_q-1,l_r-1)}^\ast),\\ 
    \end{split}
\end{equation*}
\[
\mathbf{\xi}^\ast = (\mathbf{\xi}_0^\ast,\mathbf{\xi}_1^\ast,\mathbf{\xi}_2^\ast,\dots).
\]
From the existing theory of Markov chain, $\mathbf{\xi}^\ast$ is the unique solution of the following equations.
\begin{equation}
\label{exxi}
    \mathbf{\xi}^\ast  \mathbf{Q}^{(L)} = \mathbf{0},
\end{equation}
\begin{equation}
\label{normalxi}
    \mathbf{\xi}^\ast  \mathbf{e}= 1.
\end{equation}
We can rewrite (\ref{exxi}) as
\begin{equation}
\label{mateqxi1}
\mathbf{\xi}_0^\ast \mathbf{B}^{(L)}_0 + \mathbf{\xi}_1^\ast \mathbf{A}^{(L)}_{-1} = \mathbf{0},
\end{equation}
\begin{equation}
\label{mateqxi2}
\mathbf{\xi}_{i-1}^\ast \mathbf{A}^{(L)}_1+\mathbf{\xi}_{i}^\ast \mathbf{A}^{(L)}_0  + \mathbf{\xi}_{i+1}^\ast \mathbf{A}^{(L)}_{-1} = \mathbf{0},\qquad i = 1,2,\dots.
\end{equation}
According to e.g., \cite{adan2017analysis}, $\mathbf{\xi}_{i}^\ast$ can be expressed as
\begin{equation}
\label{recR}
\mathbf{\xi}_{i}^\ast = \mathbf{\xi}_{1}^\ast \mathbf{R}^{(L)i-1}, \quad i>1,
\end{equation}
where $\mathbf{R}^{(L)}$ is the minimal non-negative solution of the following equation:
\[
\mathbf{A}^{(L)}_{1} + \mathbf{R}^{(L)}\mathbf{A}^{(L)}_{0}+ \mathbf{R}^{(L)2} \mathbf{A}^{(L)}_{-1} =\mathbf{0}.
\]
We can obtain $\mathbf{R}^{(L)}$ numerically using (\ref{recqbd}) started with $\mathbf{R}^{(L)}_0=\mathbf{O}$. $\mathbf{R}^{(L)}$ is determined recursively by repeatedly using (\ref{recqbd}) 
until $\|\mathbf{R}^{(L)}_{n+1} -\mathbf{R}^{(L)}_{n}  \|_{max} < \epsilon$ is satisfied (here, $\epsilon$ is an extremely small value and $\|\cdot\|_{max}$ denotes the max norm, i.e., the maximum absolute value of all elements of the matrix). 

\begin{equation}
\label{recqbd}
    \mathbf{R}^{(L)}_{n+1} = - (\mathbf{A}^{(L)}_{1}  + \mathbf{R}^{(L)2}_n \mathbf{A}^{(L)}_{-1})\mathbf{A}^{(L)-1}_{0}.
\end{equation}
Then, we can determine $\mathbf{\xi}_0^\ast$ and $\mathbf{\xi}_1^\ast$ by solving
\begin{equation}
    (\mathbf{\xi}_{0}^\ast, \mathbf{\xi}_{1}^\ast  ) 
    \begin{pmatrix}
    \mathbf{B}^{(L)}_{0}& \mathbf{A}^{(L)}_{1} \\
    \mathbf{A}^{(L)}_{-1} & \mathbf{A}^{(L)}_{0} + \mathbf{R}^{(L)}\mathbf{A}^{(L)}_{-1}
    \end{pmatrix}
    = (\mathbf{0}, \mathbf{0}),
\end{equation}
and the rewritten form of normalization condition (\ref{normalxi}) as
\begin{equation}
    \mathbf{\xi}_{0}^\ast \mathbf{e} + \mathbf{\xi}_{1}^\ast (\mathbf{I}-\mathbf{R}^{(L)})^{-1}\mathbf{e} = 1,
\end{equation}
where $\mathbf{I}$ is an identity matrix of appropriate size.
We also obtain the values of $\mathbf{\xi}_i^\ast$ $(i=2,3,\dots)$ by using (\ref{recR}) recursively.

\section{Analysis of GI/M/1-type Markov chain for waiting bus customers}
\label{GIM1}
The infinitesimal generator $\mathbf{Q}^{(N)}$ is given as follows:
\[\mathbf{Q}^{(N)}= \bordermatrix{
& 0 &1 &2  & \cdots & C_{bus} & \cdots &  \cdots & \cdots 	 \cr
0 & \mathbf{B}^{(N)}_0& \mathbf{A}^{(N)}_0 & \mathbf{O} & \cdots & \mathbf{O}  &  \mathbf{O} &  \mathbf{O}  & \cdots 	 \cr
1 & \mathbf{B}^{(N)}_1& \mathbf{A}^{(N)}_1 & \mathbf{A}^{(N)}_0  & \cdots & \mathbf{O}  &  \mathbf{O} &  \mathbf{O}  & \cdots 	  \cr
2 & \mathbf{B}^{(N)}_1 & \mathbf{O} & \mathbf{A}^{(N)}_1 & \cdots  & \mathbf{O}  &  \mathbf{O} &  \mathbf{O}  & \cdots 	 \cr
\vdots & \vdots  & \vdots   & \vdots  &  \ddots & \vdots  & \vdots   & \vdots  &  \ddots  \cr
C_{bus}& \mathbf{B}^{(N)}_1 & \mathbf{O} &\mathbf{O} & \cdots & \mathbf{A}^{(N)}_1 & \mathbf{A}^{(N)}_0  &  \mathbf{O} 	& \cdots   \cr
\vdots & \mathbf{O} & \mathbf{B}^{(N)}_1  & \mathbf{O}  & \cdots & \mathbf{O}& \mathbf{A}^{(N)}_1&  \mathbf{A}^{(N)}_0& \cdots	  \cr
\vdots & \mathbf{O} & \mathbf{O} &  \mathbf{B}^{(N)}_1   & \cdots &  \mathbf{O}&  \mathbf{O}&  \mathbf{A}^{(N)}_1& \cdots	  \cr
\vdots & \vdots  & \vdots   & \ddots  &  \ddots &  \ddots&  \ddots&  \ddots&  \ddots \cr
},\]
where
\[\mathbf{A}^{(N)}_0= (1-p)\lambda_n \mathbf{I}
,\]
\[\mathbf{A}^{(N)}_1=-\{(1-p)\lambda_n+l_{r}/b_n \}\mathbf{I} + 
    \bordermatrix{
    &0& 1 & \cdots& l_r-2 & l_r-1 \cr\\
    0& 0  & l_{r}/b_n& \cdots& 0& 0  \cr\\
    1& 0 &0 &   \ddots& 0 & 0 \cr\\
    \vdots& \vdots & \ddots& \ddots&  \ddots& 0 \cr\\
     l_r-2& 0 & 0&   0& 0 &l_{r}/b_n\cr\\
    l_r-1& 0 & 0&   0& 0 & 0 \cr
    }
,\]
\[\mathbf{B}^{(N)}_0=-\{(1-p)\lambda_n+l_{r}/b_n \}\mathbf{I} + 
    \bordermatrix{
    &0& 1 & \cdots& l_r-2 & l_r-1 \cr\\
    0& 0  & l_{r}/b_n& \cdots& 0& 0  \cr\\
    1& 0 &0 &   \ddots& 0 & 0 \cr\\
    \vdots& \vdots & \ddots& \ddots&  \ddots& 0 \cr\\
     l_r-2& 0 & 0&   0& 0 &l_{r}/b_n\cr\\
    l_r-1& l_{r}/b_n & 0&   0& 0 & 0 \cr
    }
,\]
\[\mathbf{B}^{(N)}_1=
    \bordermatrix{
    &0& 1 & 2& \cdots& l_r-1 \cr\\
    0& 0 & 0& 0& 0& 0 \cr\\
    1& 0 & 0& 0& 0& 0 \cr\\
    \vdots& \vdots & \vdots&  \vdots&  \vdots&  \vdots \cr\\
    l_r-2& 0 & 0& 0& 0& 0 \cr\\
    l_r-1& l_{r}/b_n & 0&  0& 0& 0 \cr
    }
.\]
First, we derive the stability condition. The infinitesimal generator $\mathbf{A}^{(N)}$ of the phase is given as
\[
\mathbf{A}^{(N)} = \mathbf{B}^{(N)}_{1}+ \mathbf{A}^{(N)}_{1}+ \mathbf{A}^{(N)}_{0}.
\]
We assume that the steady state probability of the phase is 
\[
\mathbf{\psi} = (\psi_0,\psi_1,\psi_2,\dots,\psi_{l_r-1}),
\]
which satisfies
\[
\mathbf{\psi} \mathbf{A}^{(N)} = \mathbf{0},
\]
\[
\mathbf{\psi} \mathbf{e} = 1.
\]
We can easily find the solution of $\mathbf{\psi}$ as
\[
\mathbf{\psi} =\dfrac{1}{l_r}\mathbf{e}^\top.
\]
According to e.g., \cite{adan2017analysis}, the stability condition can be written as
\begin{gather*}
 \mathbf{\psi} \mathbf{A}^{(L)}_{0}\mathbf{e}  -C_{bus}\mathbf{\psi} \mathbf{B}^{(L)}_{1}\mathbf{e} < 0 \notag \\
\Longrightarrow \lambda_{n}  (1-p) <  \dfrac{C_{bus}}{b_{n}},
\end{gather*}
which is equivalent to (\ref{busstability}).

Under the stability condition, we can derive the stability condition of our Markov chain, i.e., $\omega_{j,r}^\ast $, as follows. 
To this end, we define $\mathbf{\omega}_{j}^\ast$ and $\mathbf{\omega}^\ast$ as
\[
\mathbf{\omega}_{j}^\ast = (\omega_{j,0},\omega_{j,1},\omega_{j,2},\dots,\omega_{j,l_r-1}),
\]
\[
\mathbf{\omega}^\ast = (\mathbf{\omega}_{0}^\ast,\mathbf{\omega}_{1}^\ast,\mathbf{\omega}_{2}^\ast,\dots).
\]
We have the equilibrium equations and the probability normalization condition as follows:
\begin{equation}
\label{balanceomega}
\mathbf{\omega}^\ast \mathbf{Q}^{(N)} = \mathbf{0},
\end{equation}
\begin{equation}
\label{normalizeomega}
\mathbf{\omega}^\ast \mathbf{e} = 1.
\end{equation}
$\mathbf{\omega}_{j}^\ast $ $(j \geqq 1)$ can be written in matrix-geometric form:
\begin{equation}
\label{ngeomrtric}
\mathbf{\omega}_{j}^\ast = \mathbf{\omega}_{1}^\ast \mathbf{R}^{(N)j-1}, \qquad j\geqq 1.
\end{equation}
The balance equation (\ref{balanceomega}) can be rewritten as
\begin{equation}
\label{j0}
\mathbf{\omega}_{0}^\ast\mathbf{B}^{(N)}_0 +\sum_{j=1}^{C_{bus}} \mathbf{\omega}_{i}^\ast\mathbf{B}^{(N)}_1 = \mathbf{0},
\end{equation}
\begin{equation}
\label{jnot0}
\mathbf{\omega}_{j-1}^\ast\mathbf{A}^{(N)}_0+\mathbf{\omega}_{j}^\ast\mathbf{A}^{(N)}_1+\mathbf{\omega}_{j+C_{bus}}^\ast\mathbf{B}^{(N)}_1 = \mathbf{0},\qquad j \geqq 1.
\end{equation}
Rewriting (\ref{jnot0}) using (\ref{ngeomrtric}), we obtain
\begin{equation}
    \label{RN}
    \mathbf{A}^{(N)}_0+ \mathbf{R}^{(N)}\mathbf{A}^{(N)}_1+ \mathbf{R}^{(N)C_{bus}+1}\mathbf{B}^{(N)}_1 = \mathbf{0}.
\end{equation}
According to the existing theory, $\mathbf{R}^{(N)} $ is minimal non-negative solution of (\ref{RN}), and its numerical value can be obtained same as $\mathbf{R}^{(L)}$, i.e., by repeatedly using the following relationship until $\|\mathbf{R}^{(N)}_{n+1} -\mathbf{R}^{(N)}_{n}  \|_{max} < \epsilon$ is satisfied:
\[
\mathbf{R}^{(N)}_{n+1} = -(\mathbf{A}^{(N)}_0+\mathbf{R}_n^{(N)C_{bus}+1}\mathbf{B}^{(N)}_1 )\mathbf{A}^{(N)_1-1}.
\]
By rewriting (\ref{j0}) and (\ref{jnot0}), we can obtain the following equation to determine the unknown quantities $\mathbf{\omega}_{0}^\ast$ and $\mathbf{\omega}_{1}^\ast$:
\begin{equation}
\label{omega1}
    \mathbf{\omega}_{0}^\ast\mathbf{B}^{(N)}_0 +\sum_{j=1}^{C_{bus}}  \mathbf{\omega}_{1}^\ast \mathbf{R}^{(N)j-1}\mathbf{B}^{(N)}_1 = \mathbf{0}
\end{equation}
\begin{equation}
    \mathbf{\omega}_{0}^\ast\mathbf{A}^{(N)}_0+\mathbf{\omega}_{1}^\ast\mathbf{A}^{(N)}_1 + \mathbf{\omega}_{1}^\ast \mathbf{R}^{(N)C_{bus}}\mathbf{B}^{(N)}_1 = \mathbf{0}
\end{equation}
Besides, by rewriting the normalization condition (\ref{normalizeomega}), we obtain
\begin{equation}
\label{omega2}
    \mathbf{\omega}_{0}^\ast \mathbf{e}+ \mathbf{\omega}_{1}^\ast (\mathbf{I}-\mathbf{R}^{(N)})^{-1} \mathbf{e} = 1.
\end{equation}
By solving the simultaneous equation (\ref{omega1}) -- (\ref{omega2}), we obtain the values of $\mathbf{\omega}_{0}^\ast$ and $\mathbf{\omega}_{1}^\ast$, and we can determine $\mathbf{\omega}_{j}^\ast$ $(j \geqq 2)$ using the geometric form (\ref{ngeomrtric}).

\pagebreak
\section{PnR Monte Carlo Simulation}
\label{MCsim}
\begin{algorithm}
\caption{PNR Simulation}
\begin{algorithmic}[1]
\Procedure{PNRSim}{$\lambda, s_n, d_n, C, b, p, \lambda_c, ET_c, k_{\text{exp}}$}
    \State $k \gets \text{ComputeK}(s_n, d_n, \lambda_c, ET_c, k_{\text{exp}})$
    \State $\mu \gets s_n \times k$
    \State \text{StabilityCheck}($\lambda$, $\mu$, C, b, p)
   
    \State $t_{\lambda}, t_b, t_{(1/\mu)} \gets 0, b, 1/\mu$
    \State $w, state, time, i, j \gets 0, 0, 0, 0, 0$
    \State $data \gets \text{DataFrame}$
    \State $v, b \gets [], []$
    
    \For{$state \gets 1$ to $N$}
        \State $t_{\lambda} \gets \lambda \sim \text{ExponentialDistribution}$  
        \State $i \gets \min(t_{\lambda}, t_b, t_{(1\mu)})$
            
        \If{$i = t_{\lambda}$} 
            \State $\triangleright$ Customer arrival at PnR
            \State $data[\text{Arrival at PnR}][j] \gets time + i$
            \If{$p \leq r \sim \text{UniformDistribution}$} 
                \State $data[\text{Mode of Transport}][j] \gets \text{'car'}$
                \State append $j$ to $v$
                \State $data[\text{Departure from PnR}][j] \gets time + i$
            \Else 
                \State $w \mathrel{+}= 1$
                \State $data[\text{Mode of Transport}][j] \gets \text{'bus'}$
                \State append $j$ to $b$
            \EndIf
            \State $j \mathrel{+}= 1$
            \State $t_b \mathrel{-}= i$
            \State $t_{(1/\mu)} \mathrel{-}= i$
            \algstore{bkbreak}
\end{algorithmic}
\end{algorithm}

\begin{algorithm}
\caption{PNR Simulation - continued}
\begin{algorithmic}[1]
\algrestore{bkbreak}
        \ElsIf{$i = t_b$}
            \State $\triangleright$ Bus arrival at PnR
            \If{$b = []$} 
                \State bus departs empty
            \Else
                \For{$\beta \gets$ first $\min(w, C)$ elements of $b$}
                    \State passenger $\beta$ enters the bus
                \EndFor
                \State $data[\text{Departure from PnR}][j] \gets time + i$
                \State \textbf{delete} first $\min(w, C)$ elements of $b$
                \State $w \mathrel{-}= \min(w, C)$
            \EndIf
            \State append $j$ to $v$ 
            \State $t_b \gets b$
            \State $t_{(1/\mu)} \mathrel{-}= i$
        \ElsIf{$i = t_{(1/\mu)}$}
            \State $\triangleright$ Vehicle arrives at service station
            \For{$\nu \gets$ elements of $v$}
                \State $data[\text{Departure from SS}][\nu] \gets time + i$ \Comment{SS: Service Station}
                \State $t_{service} \gets data[\text{Departure from SS}][\nu] - data[\text{Departure from PnR}][\nu]$
                \State $data[\text{Speed}][\nu] \gets 1 / (k \times t_{service})$
                \State $\triangleright$ Calculate emission
                \State $\tau \gets data[\text{Mode of Transport}][\nu]$
                \State $speed \gets data[\text{Speed}][\nu]$
                \If{$\tau = \text{'bus'}$}
                    \State $data[\text{Bus Emission}][\nu] \gets \text{MEET}_{bus}(speed, C)$
                \Else
                    \State $fuel \gets f \sim \text{Bernoulli}(p_{gasoline})$
                    \If{$f = 0$} 
                        \State $data[\text{Car Emission}][\nu] \gets \text{MEET}_{car}^{diesel}(speed)$
                    \ElsIf{$f = 1$} 
                        \State $data[\text{Car Emission}][\nu] \gets \text{MEET}_{car}^{gasoline}(speed)$
                    \EndIf
                \EndIf
            \EndFor
            \State update $v$
            \State $t_b \mathrel{-}= i$
            \State $t_{(1/\mu)} \gets 1 / \mu$
        \EndIf
    \EndFor
\EndProcedure
\end{algorithmic}
\end{algorithm}

\pagebreak
\section{Supplementary figures} 
\label{SUPfigs}
\begin{figure}[ht]
\centering
\subfloat[hub 1 emission][Total car emissions for Hub 1]{
\includegraphics[width=0.45\textwidth]{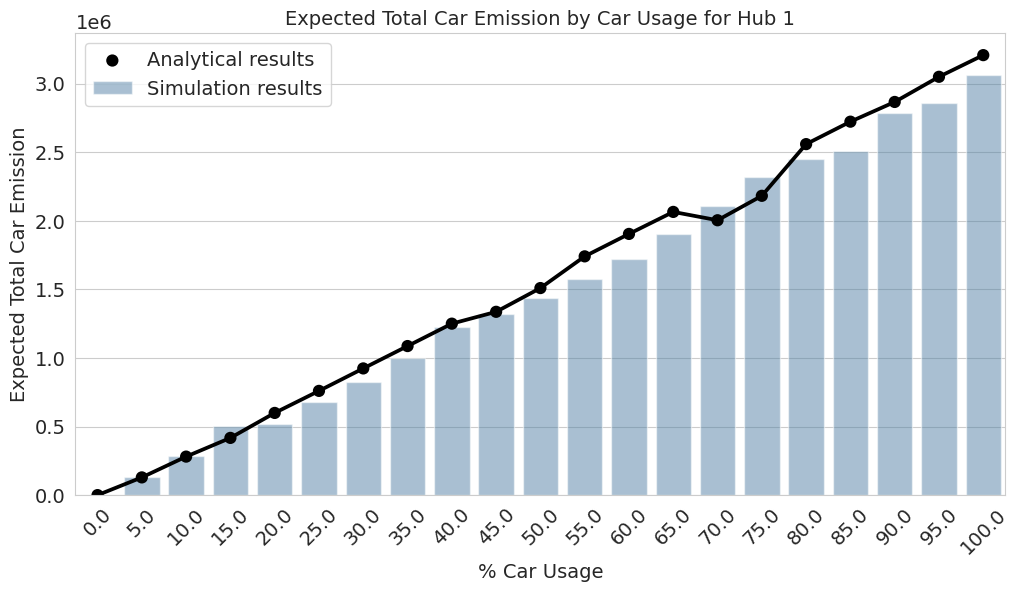}
\label{fig:hub1ce}}
\subfloat[hub 1 emission][Total bus emissions for Hub 1]{
\includegraphics[width=0.45\textwidth]{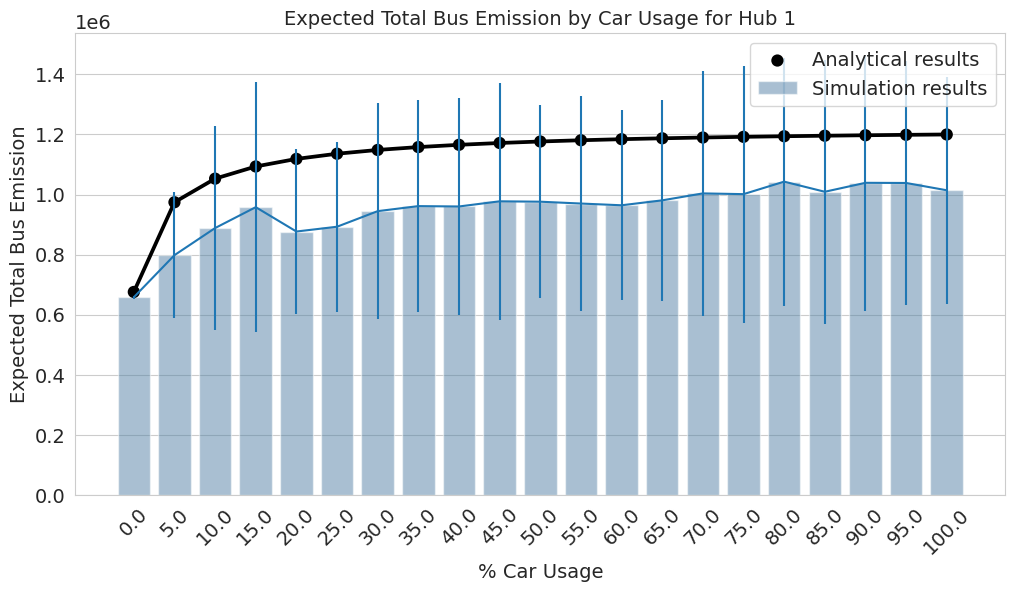}
\label{fig:hub1be}}

\subfloat[hub 2 emission][Total car emissions for Hub 2]{
\includegraphics[width=0.45\textwidth]{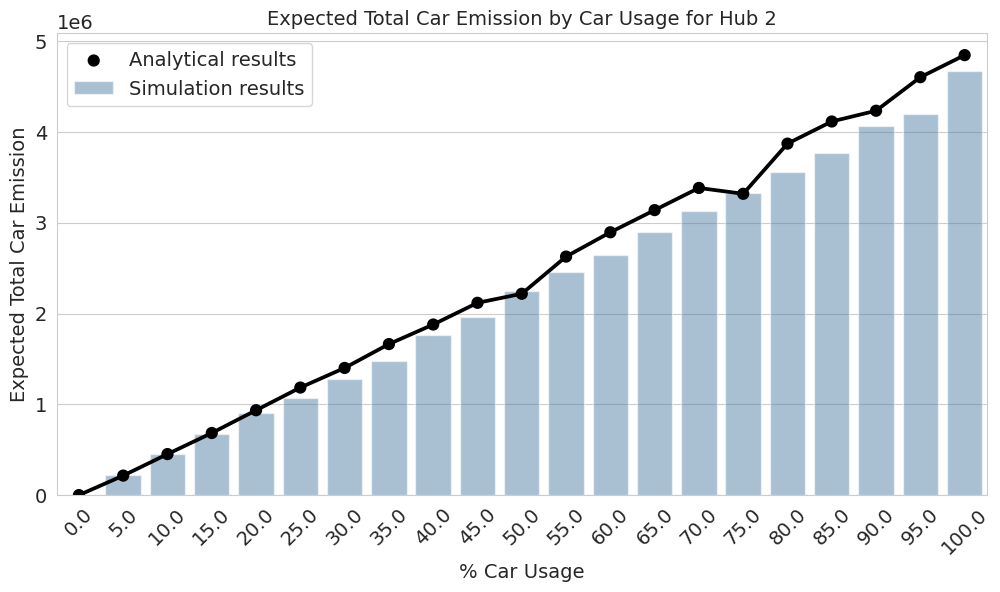}
\label{fig:hub2ce}}
\subfloat[hub 2 emission][Total bus emissions for Hub 2]{
\includegraphics[width=0.45\textwidth]{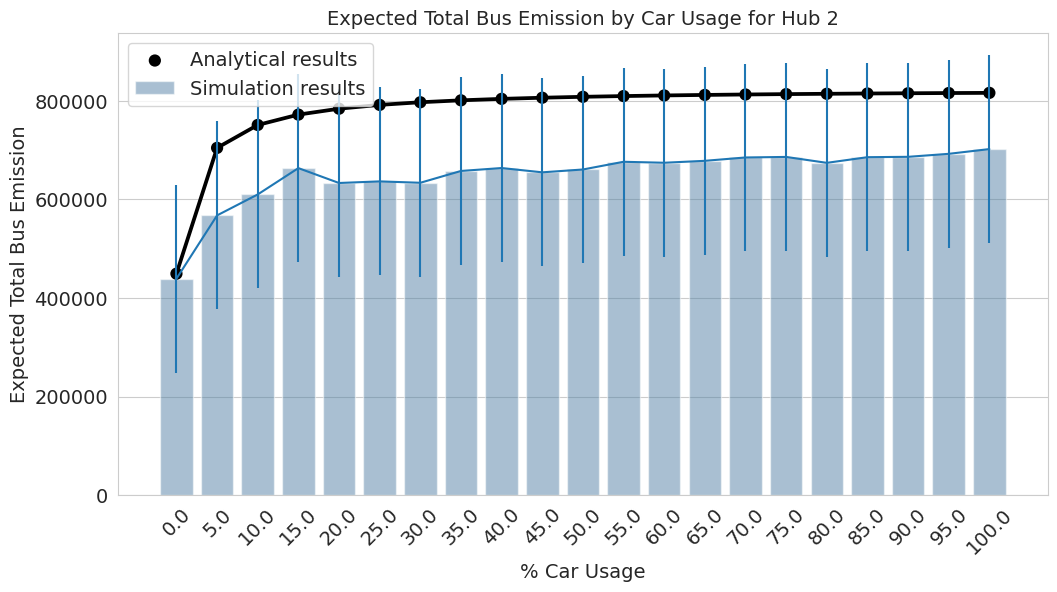}
\label{fig:hub2be}}

\subfloat[hub 3 emission][Total car emissions for Hub 3]{
\includegraphics[width=0.45\textwidth]{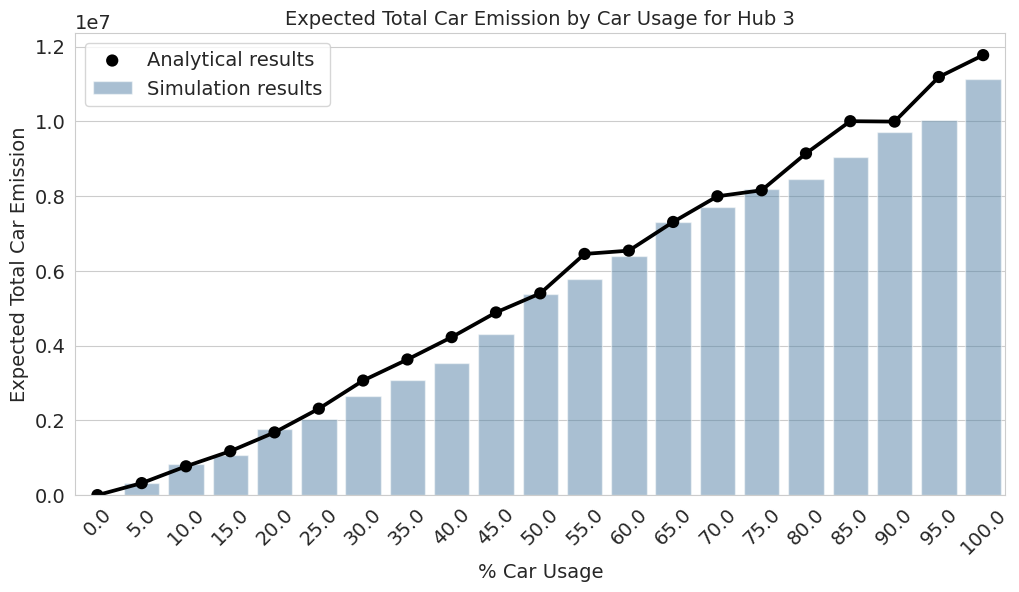}
\label{fig:hub3ce}}
\subfloat[hub 3 emission][Total bus emissions for Hub 3]{
\includegraphics[width=0.45\textwidth]{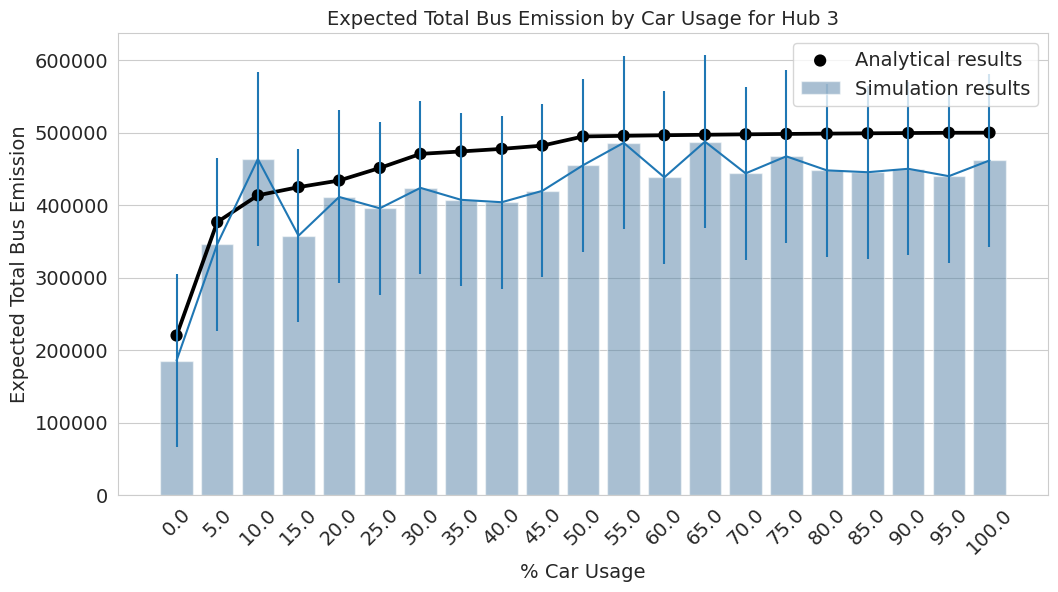}
\label{fig:hub3be}}

\caption{Total car and bus emissions in grams for all hubs. Vertical bars represent simulation results while points and line represent the approximation model's analytical results. Analytical results for the arrival of cars and buses to the service station using Erlang distribution with phase 20 and 200 respectively. For bus emissions we have included a 95\% confidence interval bar to demonstrate that analytical results fall inside the 95\% confidence interval of the simulated results.}
\label{fig:hubsve1}
\end{figure}

\begin{figure}
\ContinuedFloat
\centering
\subfloat[hub 4 emission][Total car emissions for Hub 4]{
\includegraphics[width=0.45\textwidth]{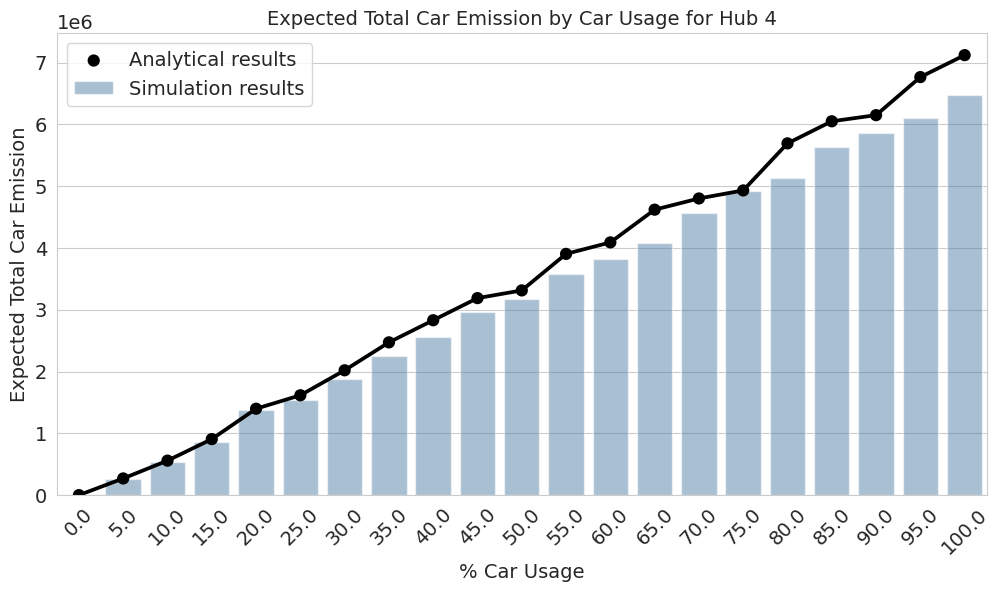}
\label{fig:hub4ce}}
\subfloat[hub 4 emission][Total bus emissions for Hub 4]{
\includegraphics[width=0.45\textwidth]{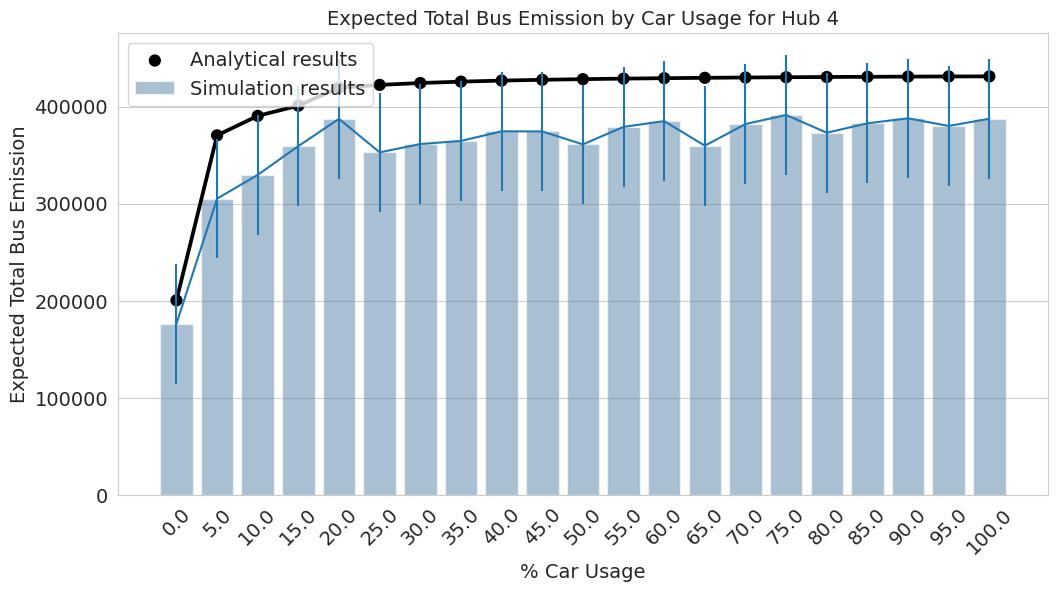}
\label{fig:hub4be}}

\subfloat[hub 5 emission][Total car emissions for Hub 5]{
\includegraphics[width=0.45\textwidth]{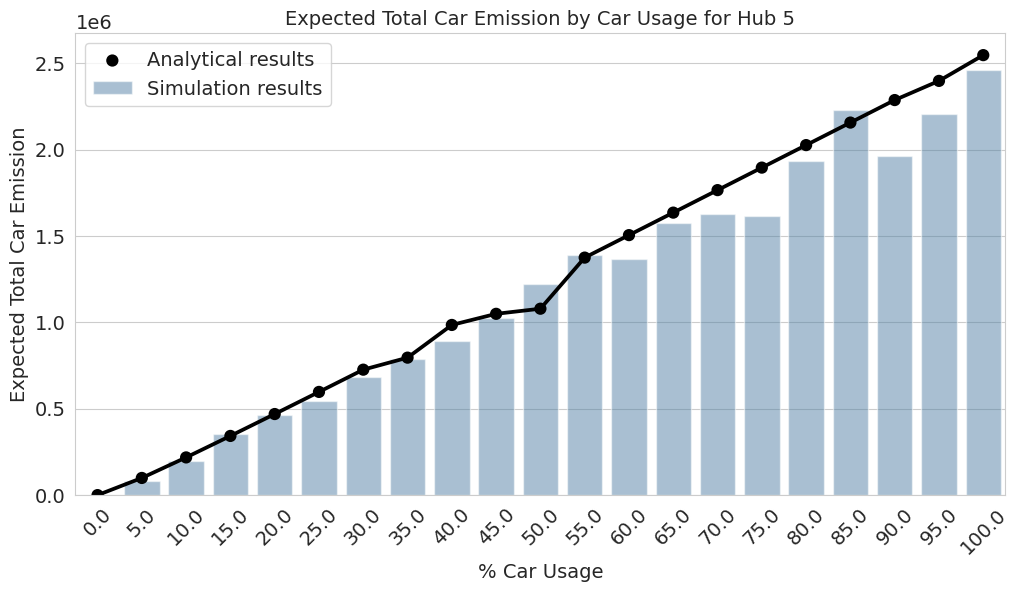}
\label{fig:hub5ce}}
\subfloat[hub 5 emission][Total bus emissions for Hub 5]{
\includegraphics[width=0.45\textwidth]{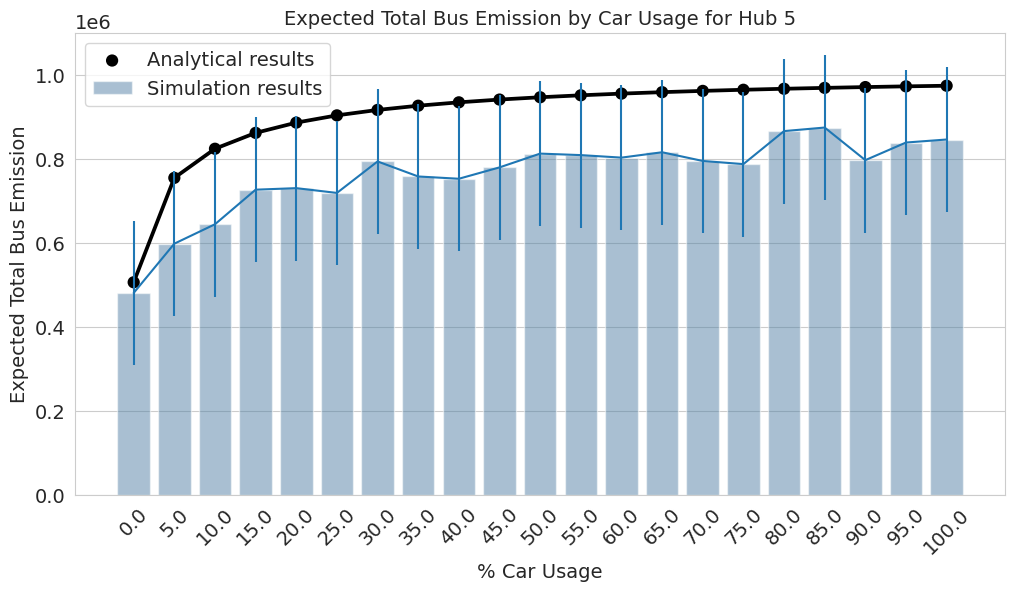}
\label{fig:hub5be}}

\caption{Total car and bus emissions in grams for all hubs. Vertical bars represent simulation results while points and line represent the approximation model's analytical results. Analytical results for the arrival of cars and buses to the service station using Erlang distribution with phase 20 and 200 respectively. For bus emissions we have included a 95\% confidence interval bar to demonstrate that analytical results fall inside the 95\% confidence interval of the simulated results.}
\label{fig:hubsve2}
\end{figure}

\begin{figure}
\centering
\subfloat[hub 1 trip times][Total trip times for Hub 1]{
\includegraphics[width=0.45\textwidth]{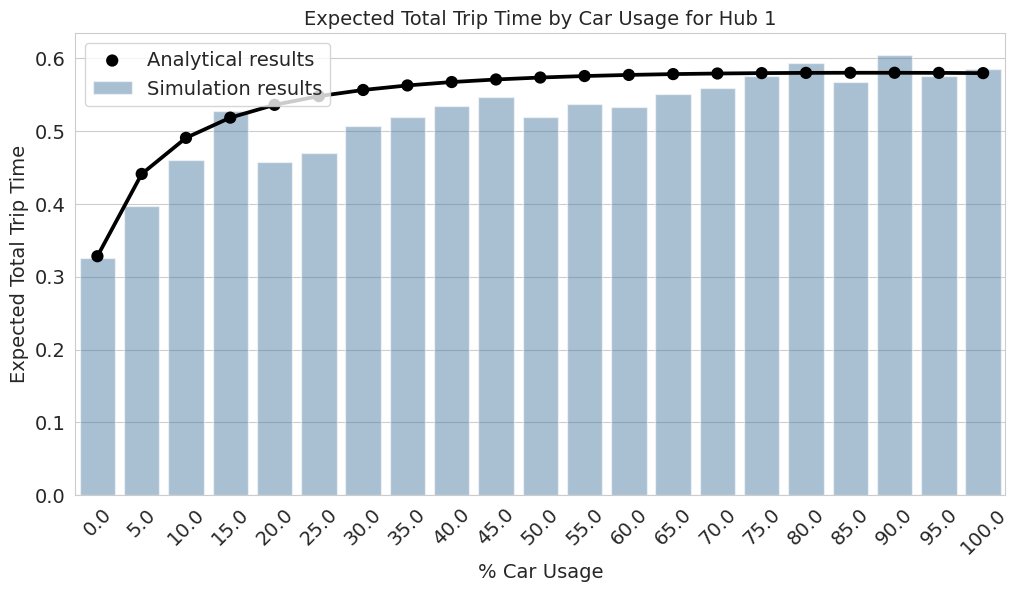}
\label{fig:hub1tt}}
\subfloat[hub 2 trip times][Total trip times for Hub 2]{
\includegraphics[width=0.45\textwidth]{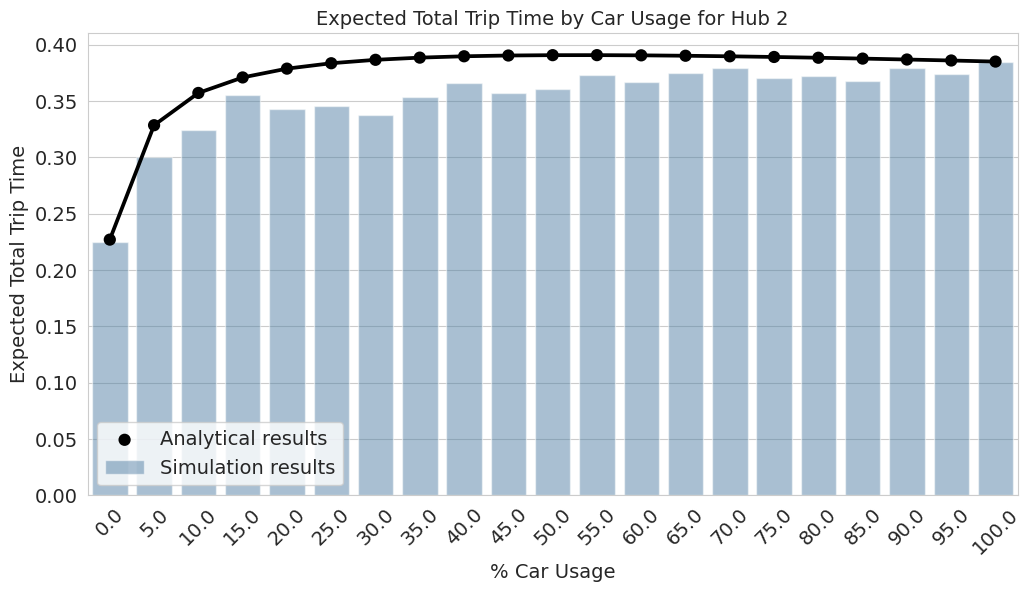}
\label{fig:hub2tt}}

\subfloat[hub 3 trip times][Total trip times for Hub 3]{
\includegraphics[width=0.45\textwidth]{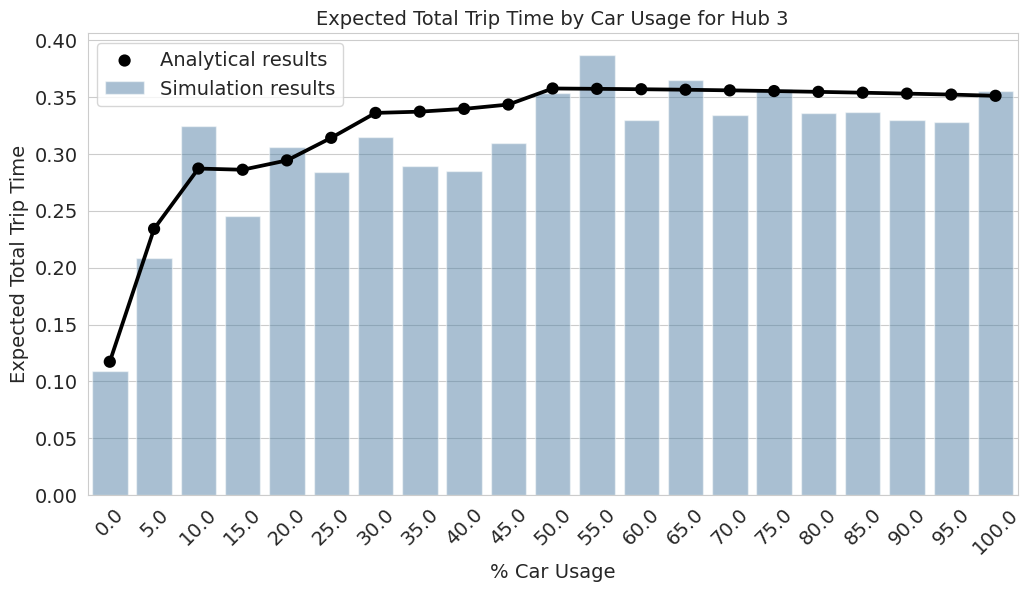}
\label{fig:hub3tt}}
\subfloat[hub 4 trip times][Total trip times for Hub 4]{
\includegraphics[width=0.45\textwidth]{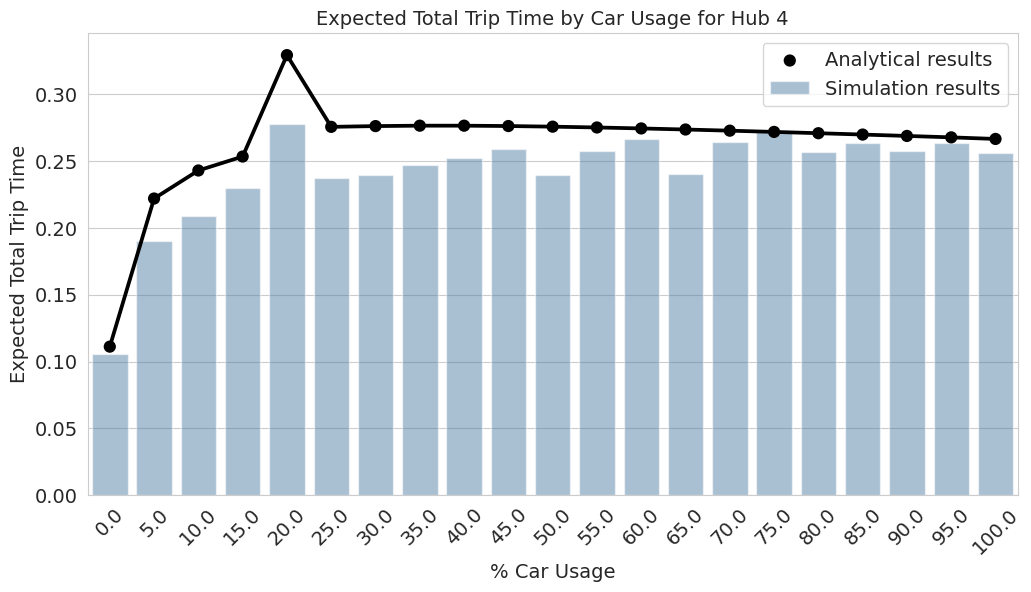}
\label{fig:hub4tt}}

\subfloat[hub 5 trip times][Total trip times for Hub 5]{
\includegraphics[width=0.45\textwidth]{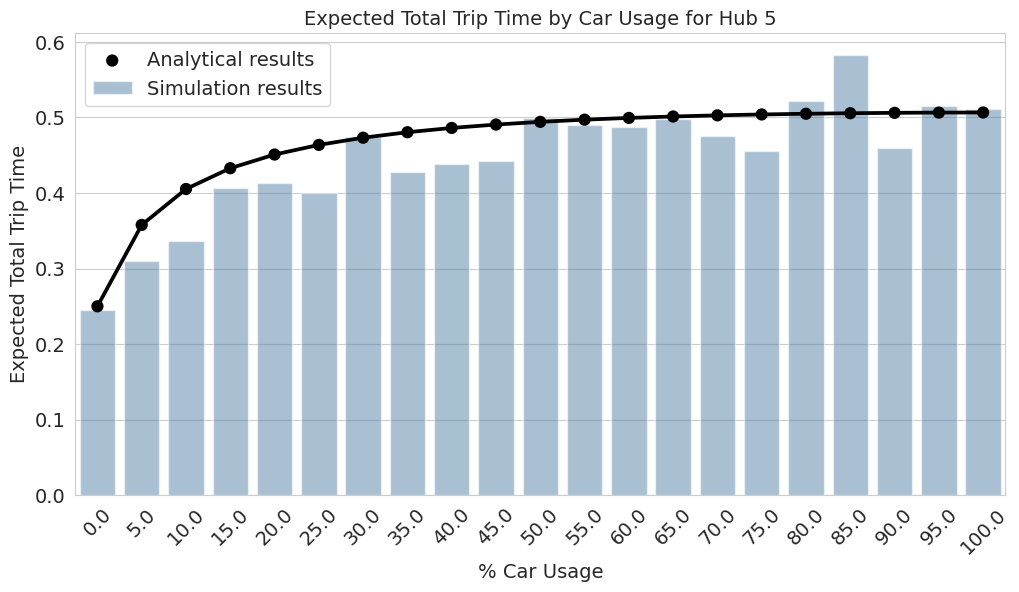}
\label{fig:hub5tt}}

\caption{Total trip time in mean hours for all hubs. Vertical bars represent simulation results while points and line represent the approximation model's analytical results.}
\label{fig:hubstimes2}
\end{figure}

\begin{figure}
\centering
\subfloat[hub 1 travel and wait times][Travel and wait times for Hub 1]{
\includegraphics[width=0.45\textwidth]{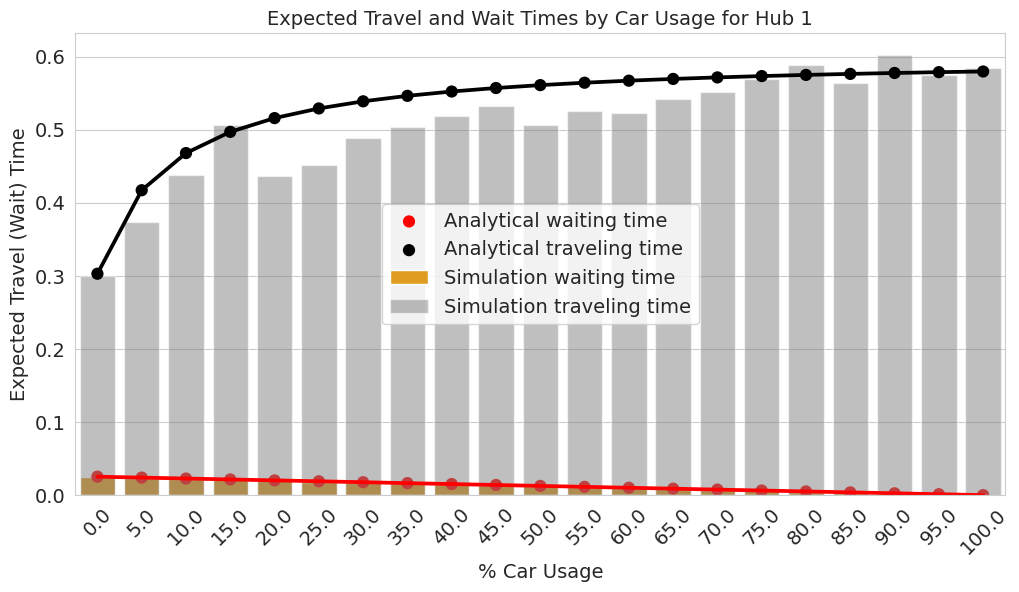}
\label{fig:hub1wt}}
\subfloat[hub 2 travel and wait times][Travel and wait for Hub 2]{
\includegraphics[width=0.45\textwidth]{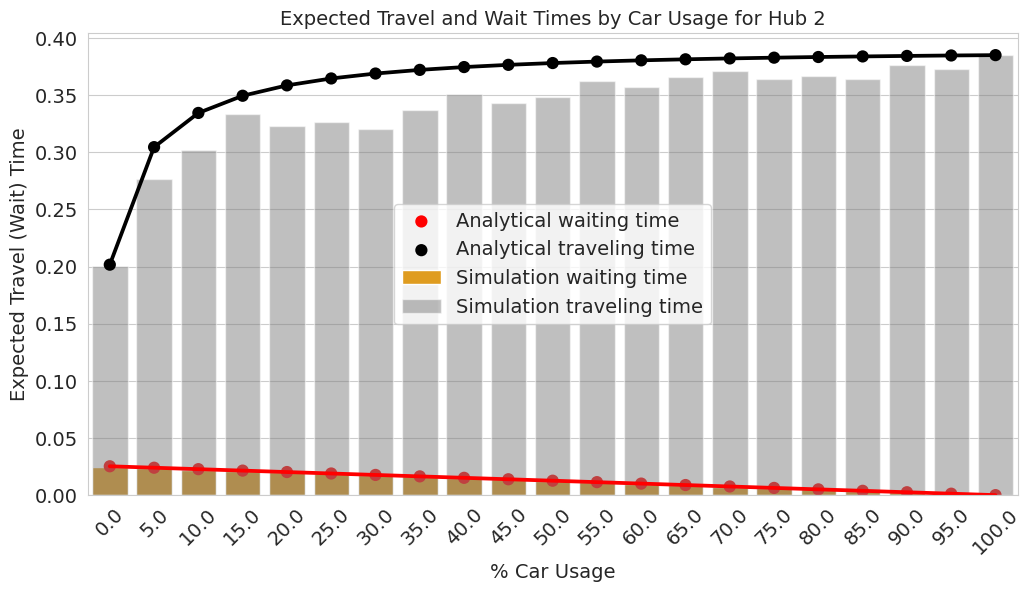}
\label{fig:hub2wt}}

\subfloat[hub 3 travel and wait times][Travel and wait for Hub 3]{
\includegraphics[width=0.45\textwidth]{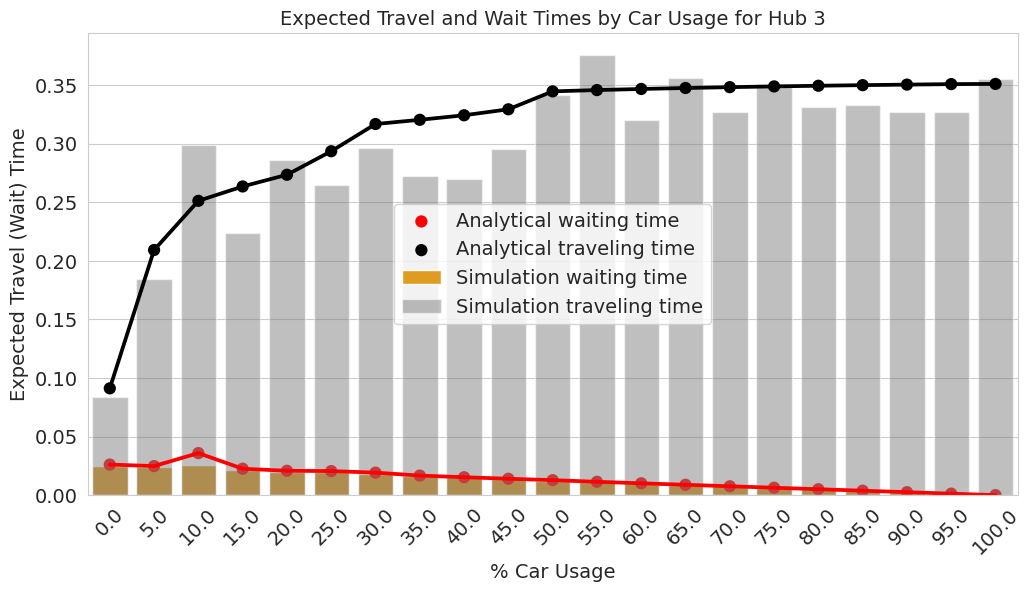}
\label{fig:hub3wt}}
\subfloat[hub 4 travel and wait times][Travel and wait for Hub 4]{
\includegraphics[width=0.45\textwidth]{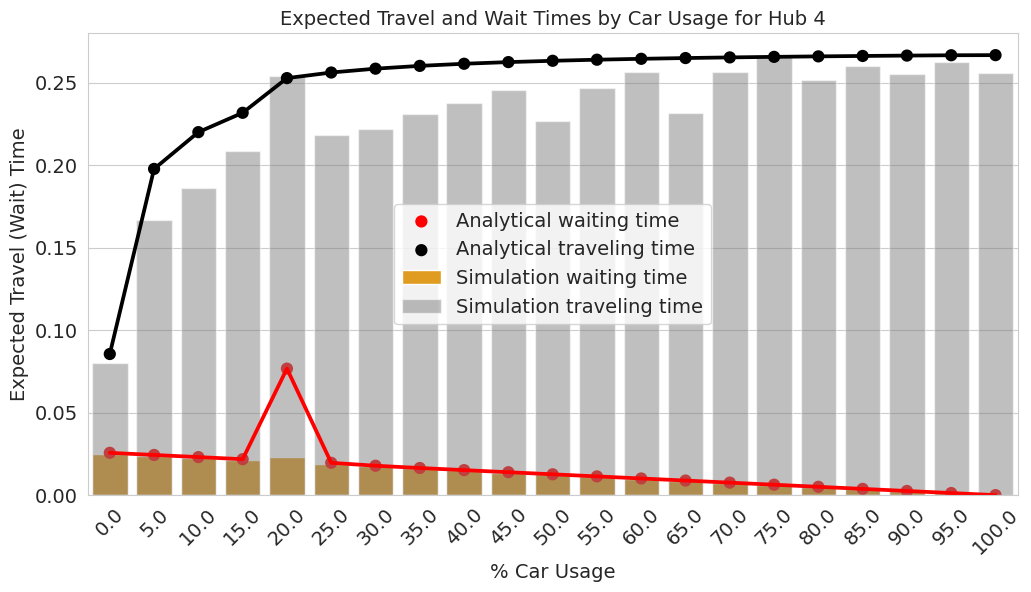}
\label{fig:hub4wt}}

\subfloat[hub 5 travel and wait times][Travel and wait for Hub 5]{
\includegraphics[width=0.45\textwidth]{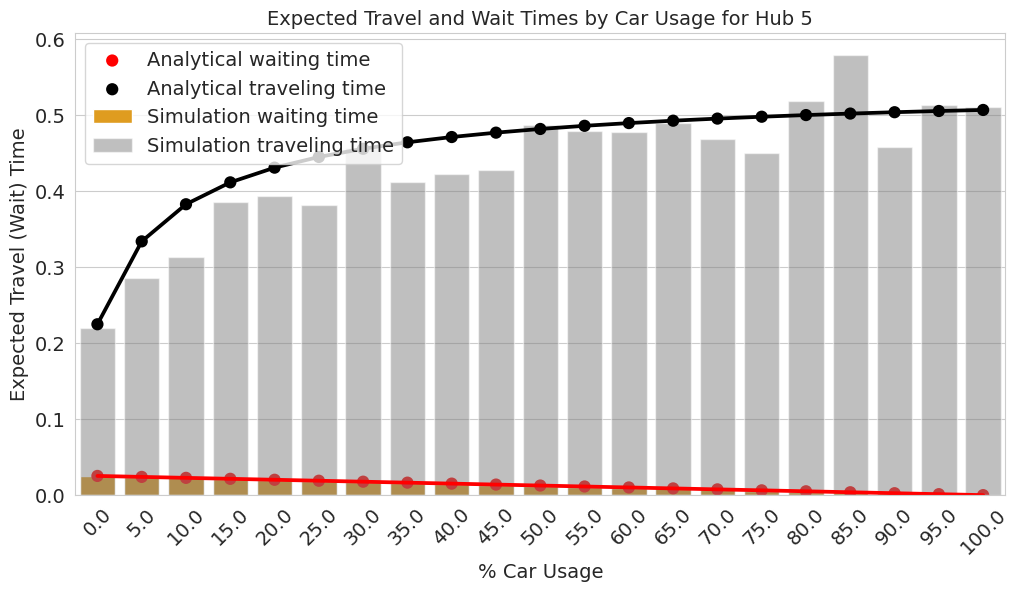}
\label{fig:hub5wt}}

\caption{Traveling and waiting times in mean hours for all hubs. Vertical bars represent simulation results while points and line represent the approximation model's analytical results.}
\label{fig:hubswtimes}
\end{figure}

\end{document}